\documentclass[twocolumn]{autart}    % Enable this line and disable the 
\usepackage{graphicx}          % Include this line if your 
\usepackage{url}

\usepackage{siunitx}
\usepackage{tikz}
\usepackage{mathdots}
\usepackage{yhmath}
\usepackage{cancel}
\usepackage{array}
\usepackage{multirow}
\usepackage{textcomp, gensymb}
\usepackage{tabularx}
\usepackage{extarrows}
\usepackage{booktabs}
\usetikzlibrary{fadings}
\usetikzlibrary{patterns}
\usetikzlibrary{shadows.blur}
\usetikzlibrary{shapes}
\usepackage{datetime}
\usepackage{wrapfig}
\usepackage[authoryear,round]{natbib}

\usepackage{cuted}
\usepackage{amsmath} % assumes amsmath package installed
\usepackage{amssymb}  % assumes amsmath package installed
\usepackage{mathtools}
\usepackage{arydshln}
\usepackage{comment}
\usepackage[normalem]{ulem}
\usepackage{xfrac}

\DeclareMathOperator*{\rank}{rank}
\DeclareMathOperator*{\diag}{diag}
\DeclareMathOperator*{\trace}{tr}

\allowdisplaybreaks

\makeatletter
\def\ps@copyright{\let\@mkboth\@gobbletwo
  \def\@oddhead{}%
  \let\@evenhead\@oddhead
  \def\@oddfoot{\small\slshape
    Article in press, \@journal\ \the\@pubyear\hfil\hbox{}}%
  \let\@evenfoot\@oddfoot
}
\makeatother

\begin{document}

\begin{frontmatter}

\title{Plant Equivalent Controller Realizations \\ for Attack-Resilient Cyber-Physical Systems\thanksref{footnoteinfo}} % Title, preferably not more 
                                                % than 10 words.

\thanks[footnoteinfo]{The research leading to these results has received funding from the European Union’s Horizon Europe programme under grant agreement No 101069748 – SELFY project, and the Ministry of Education (MOE), Singapore, under its Academic Research Fund (AcRF) Tier 1 grant, funded through the SUTD Kickstarter Initiative (SKI 2021\_06\_02).}

\author[Eindhoven]{Mischa Huisman}\ead{m.r.huisman@tue.nl},    % Add the 
\author[Eindhoven]{Erjen Lefeber}\ead{a.a.j.lefeber@tue.nl},  % (ead) as shown
\author[Eindhoven]{Nathan van de Wouw}\ead{n.v.d.wouw@tue.nl},  % (ead) as shown
\author[Eindhoven,Singapore]{Carlos Murguia}\ead{c.g.murguia@tue.nl,murguia\_rendon@sutd.edu.sg}   % e-mail address 

\address[Eindhoven]{Eindhoven University of Technology, The Netherlands}  % Please supply 
\address[Singapore]{Singapore University of Technology and Design}
%\address[Rome]{Senate House, Rome}             % full addresses
%\address[Baiae]{The White House, Baiae}        % here.

                             % keyword list or with the 
                                          % help of the Automatica 
                                          % keyword wizard

\begin{abstract}                          % Abstract of not more than 200 words.
As cyber-physical systems (CPSs) become more dependent on data and communication networks, their vulnerability to false data injection (FDI) attacks has raised significant concerns. Among these, stealthy attacks, those that evade conventional detection mechanisms, pose a critical threat to closed-loop performance. This paper introduces a controller-oriented method to enhance CPS resiliency against such attacks without compromising nominal closed-loop behavior. Specifically, we propose the concept of plant equivalent controller (PEC) realizations, representing a class of dynamic output-feedback controllers that preserve the input-output behavior of a given base controller while exhibiting distinct robustness properties in the presence of disturbances and sensor attacks. To quantify and improve robustness, we employ reachable set analysis to assess the impact of stealthy attacks on the closed-loop dynamics. Building on this analysis, we provide mathematical tools (in terms of linear matrix inequalities) to synthesize the optimal PEC realization that minimizes the reachable set under peak-bounded disturbances. The proposed framework thus provides systematic analysis and synthesis tools to enhance the attack resilience of CPSs while maintaining the desired nominal performance. The effectiveness of the approach is demonstrated on the quadruple-tank process subject to stealthy sensor attacks.
\end{abstract}

\end{frontmatter}

%% Introduction %%
\section{Introduction}
Cyber-physical systems (CPSs) constitute a class of networked control systems with diverse and promising applications, including power grids~\citep{peng_survey_2019}, autonomous vehicles~\citep{sun_survey_2022}, and water distribution systems~\citep{amin_cyber_2013}. However, the increasing reliance on data and communication networks also increases the vulnerability of CPSs to malicious cyberattacks, particularly false data injection (FDI) attacks, which could negatively affect the physical domain of the CPS. This has motivated extensive research on CPS security, intending to develop technologies that enhance system resiliency~\citep{dibaji_systems_2019,ferrari_safety_2021}. 

In general, resiliency can be enhanced through three primary mechanisms: prevention, detection, and mitigation~\citep{teixeira_secure_2015}. Prevention and detection methods have been widely studied, where various methods have been proposed to strengthen CPS resilience~\citep{giraldo_survey_2019,musleh_survey_2020,ju_survey_2022,chowdhury_attacks_2020,nguyen_security_2025_1}. Nevertheless, their effectiveness is limited by unknown process and measurement disturbances~\citep{musleh_survey_2020,teixeira_attack_2012,zhang_networked_2019}, and by adversaries exploiting system model knowledge~\citep{anand_risk_2022,teixeira_strategic_2015}, enabling the design of stealthy attacks that can evade even the most advanced detection schemes~\citep{chong_tutorial_2019,ding_secure_2021}.

From a control-theoretic perspective, two main formulations have been proposed in the literature to mitigate the effect of FDI attacks and enhance the robustness of CPSs: (i) \emph{active} methods that employ fallback or switching control strategies, and (ii) \emph{passive} attack-resilient methods that seek to withstand the effect of stealthy attacks. Active methods are of significant interest~\citep{rodriguez_arozamena_fault_2025,gheitasi_safety_2021,su_static_2020,ghaderi_blended_2021,franze_resilient_2022,arauz_cyber_security_2022,zhu_game_theoretic_2015}, where controllers, combined with anomaly detection schemes or attacker models, restrict the feasibility of stealthy attacks. Although interesting, the effectiveness of the control law depends on the detector's performance or the accuracy of the attack model, and, therefore, may not inherently improve robustness when the attacker remains stealthy. Passive attack-resilient methods include: optimal allocation of security measures~\citep{anand_risk_2022}, control input filtering~\citep{escudero_safety_preserving_2023,attar_datadriven_2025}, safety controllers~\citep{gheitasi_safety_2021,gheitasi_worst_case_2022,lin_secondary_2023}, and saturation of control signals~\citep{hadizadeh_kafash_constraining_2018}. These methods primarily focus on security without explicitly addressing the effect of these measures on closed-loop performance, or they are designed to minimize the distortion of attack-free signals.

\begin{comment}
%% ORIGINAL TEXT
From a control-theoretic perspective, two main formulations have been proposed in the literature to mitigate the effect of FDI attacks and enhance the robustness of CPSs: (i) \emph{active} methods that employ fallback or switching control strategies, and (ii) \emph{passive} attack-resilient methods that seek to withstand the effect of stealthy attacks. Although active methods are of significant interest, they rely on detection schemes to trigger fallback strategies~\citep{rodriguez-arozamena_fault_2025,gheitasi_safety_2021,su_static_2020}, and therefore do not enhance robustness when the attacker remains stealthy. Passive attack-resilient methods include: optimal allocation of security measures~\citep{anand_risk_2022}, control input filtering~\citep{escudero_safety-preserving_2023,attar_datadriven_2025}, safety controllers~\citep{gheitasi_worst-case_2022,lin_secondary_2023}, and saturation of control signals~\citep{hadizadeh_kafash_constraining_2018}. However, these methods primarily focus on security without explicitly addressing the effect of these measures on closed-loop performance, or they are designed to minimize the distortion of attack-free signals.
\end{comment}

In this manuscript, we address the problem of improving the resiliency of CPSs against stealthy attacks without compromising the desired closed-loop performance. To this end, we introduce a passive, controller-oriented approach named \emph{plant equivalent controller (PEC) realizations}. By reformulating a given dynamic output-feedback controller (the \emph{base controller}), we derive a class of equivalent realizations (the \emph{PEC realizations}) that preserve the nominal input–output behavior of the base controller and therefore maintain the desired closed-loop system performance. While these realizations exhibit identical behavior in the absence of disturbances or attacks, their robustness properties differ in the presence of FDI attacks and process and measurement disturbances.

By integrating the output-feedback detector into the system formulation, the closed-loop dynamics are expressed in terms of the detector's state-estimation error, the residual, and the process and measurement disturbances. As the robustness properties of different controller realizations vary, reachable-set analysis is used to quantify the robustness of different PEC realizations against stealthy attacks. Reachable sets have been used in the literature, e.g.,~\citep{gheitasi_safety_2021,escudero_safety_preserving_2023,attar_datadriven_2025}, to evaluate robustness by characterizing the states an attacker can reach while remaining undetected. In this manuscript, we compute an ellipsoidal overapproximation of the closed-loop reachable set induced by bounded state-estimation errors, residuals, and process and measurement disturbances, assuming the attacker aims to remain stealthy. Its volume serves as the optimality criterion in a semi-definite program that computes an \emph{optimal PEC realization}. The proposed method enhances CPS robustness without compromising essential system properties such as stability or reference-tracking performance

The main contributions of this manuscript are summarized as follows: 1) We derive a class of equivalent controller realizations (\emph{PEC realizations}) that preserve the nominal closed-loop performance of a given dynamic output-feedback controller (\emph{base controller}); 2) We develop a set of semi-definite programs that compute an \emph{optimal PEC realization} by minimizing the reachable set of the overall LTI closed-loop system, assuming the attacker aims to remain stealthy. 

The remainder of the paper is organized as follows. Sections~\ref{sec:preliminaries} and~\ref{sec:SA} introduce the notation, key definitions, and problem formulation. In Section~\ref{sec:CR}, the PEC realizations are derived, and conditions ensuring their existence are established. Section~\ref{sec:DI} presents the integration of the detector scheme into the closed-loop system and the formulation to compute the optimal PEC realization. In Section~\ref{sec:RESULTS}, the proposed framework is applied to the quadruple-tank process, and a simulation study is conducted to demonstrate the performance under a stealthy attack. Finally, conclusions and recommendations are given in Section~\ref{sec:Conclusion}.

\section{Notation and Definitions}\label{sec:preliminaries}

%\subsection{Notation}
The symbol $\mathbb{R}$ represents the set of real numbers, while $\mathbb{R}_{>0}$ ($\mathbb{R}_{\geq 0}$) denotes the set of positive (non-negative) real numbers. The symbol $\mathbb{N}$ stands for the set of natural numbers, including zero. The $n \times m$ matrix composed of only zeros is denoted by $\mathbf{0}_{n \times m}$, or $\mathbf{0}$ when its dimension is clear. Consider a finite index set $\mathcal{L} := \{l_1,\ldots,l_k\}  \subset \mathbb{N}$, the notation diag[$B_j$] stands for the diagonal block matrix $\text{diag}[B_{l_1},\ldots,B_{l_{k}}]$. The notation $A \succeq 0$ (resp., $A \preceq 0$) indicates that the symmetric matrix $A$ is positive (resp., negative) semi-definite, i.e., all the eigenvalues of the symmetric matrix $A$ are positive (resp., negative) or equal to zero, whereas the notation $A \succ 0$ (resp., $A \prec 0$) indicates positive (resp., negative) definiteness, i.e., all the eigenvalues are strictly positive (resp., negative). 
%%% DEFENITION 1 (REACHABLE SET) %%%
\begin{defn}[Reachable Set]\emph{} \label{defn:reachableSet}
Consider the perturbed Linear Time-Invariant (LTI) system:\label{def1}
\begin{equation}
    \label{eq:LTI_set}
    \dot{\zeta}(t) = \mathcal{A} \zeta(t) + \sum_{i=1}^N \mathcal{B}_i \omega_i(t), \ \zeta(0) = \zeta_0
\end{equation}
with state $\zeta(t) \in \mathbb{R}^{n_\zeta}$, peak-bounded perturbation $\omega_i(t) \in \mathbb{R}^{p_i}$ satisfying $\mathcal{E}_{\omega_i} \coloneqq \{\omega_i(t) \in \mathbb{R}^{p_i} \mid \omega_i^\top(t) W_i \omega_i(t) \leq 1 \}$ for some positive definite matrix $W_i \in \mathbb{R}^{p_i \times p_i}, i = \{1,...,N\}, N\in \mathbb{N}$, and matrices $\mathcal{A} \in \mathbb{R}^{n_\zeta \times n_\zeta}$ and $\mathcal{B}_i \in \mathbb{R}^{n_\zeta \times p_i}$. The reachable set $\mathcal{R}_{\zeta_0}$(t) at time $t \in \mathbb{R}_{\geq0}$ from initial condition $\zeta(0)=\zeta_0 \in \mathbb{R}^{n_\zeta}$ is the set of states $\zeta(t)$ that satisfy the differential equation \eqref{eq:LTI_set} through all possible perturbations $\omega_i(t) \in \mathcal{E}_{\omega_i}$, i.e., 
\begin{equation}
    \mathcal{R}_{\zeta_0}(t) \! \coloneqq \!\!\left\{ \zeta(t) \left| \begin{array}{l}
        \exists\, \omega_i(s)\in \mathcal{E}_{\omega_i}, \text{ s.t.} \\ \zeta(s) \text{ \textit{solution to \eqref{eq:LTI_set},}}\,  s \in [0,t] \\
    \end{array} \right\} \right. .
\end{equation}
If $\mathcal{A}$ is Hurwitz, the infinite time reachable set is defined as $\mathcal{R}_{\zeta_0} (\infty) = \mathcal{R}_{\zeta_0}^\infty \coloneqq \lim_{t \to \infty} \mathcal{R}_{\zeta_0} (t)$.
\end{defn}
\section{Problem Formulation} \label{sec:SA}
\begin{figure}[tb]
    \centering
    \resizebox{0.85\columnwidth}{!}{\tikzset{every picture/.style={line width=0.75pt}} %set default line width to 0.75pt        

\begin{tikzpicture}[x=0.75pt,y=0.75pt,yscale=-1,xscale=1]
%uncomment if require: \path (0,300); %set diagram left start at 0, and has height of 300

%Flowchart: Alternative Process [id:dp2714544344458525] 
\draw  [line width=2.25]  (295.4,57.4) .. controls (295.4,53.53) and (298.53,50.4) .. (302.4,50.4) -- (358.4,50.4) .. controls (362.27,50.4) and (365.4,53.53) .. (365.4,57.4) -- (365.4,83.4) .. controls (365.4,87.27) and (362.27,90.4) .. (358.4,90.4) -- (302.4,90.4) .. controls (298.53,90.4) and (295.4,87.27) .. (295.4,83.4) -- cycle ;
%Flowchart: Alternative Process [id:dp9235464511624109] 
\draw  [line width=2.25]  (295.57,167.9) .. controls (295.57,164.03) and (298.7,160.9) .. (302.57,160.9) -- (358.57,160.9) .. controls (362.43,160.9) and (365.57,164.03) .. (365.57,167.9) -- (365.57,193.9) .. controls (365.57,197.77) and (362.43,200.9) .. (358.57,200.9) -- (302.57,200.9) .. controls (298.7,200.9) and (295.57,197.77) .. (295.57,193.9) -- cycle ;
%Flowchart: Alternative Process [id:dp3192274898048003] 
\draw  [line width=2.25]  (295.4,227.4) .. controls (295.4,223.53) and (298.53,220.4) .. (302.4,220.4) -- (358.4,220.4) .. controls (362.27,220.4) and (365.4,223.53) .. (365.4,227.4) -- (365.4,253.4) .. controls (365.4,257.27) and (362.27,260.4) .. (358.4,260.4) -- (302.4,260.4) .. controls (298.53,260.4) and (295.4,257.27) .. (295.4,253.4) -- cycle ;
%Shape: Cloud [id:dp0790167074031024] 
\draw  [line width=2.25]  (417.57,116.9) .. controls (416.74,112.85) and (419.47,108.83) .. (424.61,106.56) .. controls (429.74,104.29) and (436.38,104.16) .. (441.71,106.23) .. controls (443.6,103.87) and (447.05,102.24) .. (451.03,101.83) .. controls (455,101.43) and (459.03,102.29) .. (461.9,104.17) .. controls (463.5,102.03) and (466.66,100.59) .. (470.24,100.36) .. controls (473.83,100.14) and (477.34,101.16) .. (479.52,103.06) .. controls (482.42,100.79) and (487.04,99.84) .. (491.38,100.61) .. controls (495.72,101.38) and (498.99,103.74) .. (499.79,106.66) .. controls (503.35,107.31) and (506.31,108.95) .. (507.91,111.15) .. controls (509.52,113.36) and (509.6,115.92) .. (508.15,118.17) .. controls (511.65,121.2) and (512.47,125.23) .. (510.3,128.77) .. controls (508.13,132.3) and (503.3,134.81) .. (497.61,135.35) .. controls (497.57,138.66) and (494.83,141.71) .. (490.44,143.3) .. controls (486.06,144.9) and (480.71,144.8) .. (476.47,143.04) .. controls (474.66,147.02) and (469.57,149.94) .. (463.4,150.55) .. controls (457.23,151.16) and (451.09,149.35) .. (447.62,145.9) .. controls (443.37,147.6) and (438.28,148.09) .. (433.48,147.26) .. controls (428.68,146.43) and (424.59,144.34) .. (422.12,141.48) .. controls (417.78,141.82) and (413.58,140.32) .. (411.61,137.74) .. controls (409.64,135.16) and (410.31,132.04) .. (413.3,129.92) .. controls (409.43,128.41) and (407.45,125.41) .. (408.4,122.48) .. controls (409.35,119.56) and (413.02,117.37) .. (417.48,117.06) ; \draw  [line width=2.25]  (413.3,129.92) .. controls (415.13,130.64) and (417.24,130.96) .. (419.35,130.85)(422.12,141.48) .. controls (423.03,141.41) and (423.92,141.26) .. (424.77,141.04)(447.62,145.9) .. controls (446.98,145.26) and (446.45,144.58) .. (446.03,143.87)(476.47,143.04) .. controls (476.8,142.32) and (477.01,141.57) .. (477.11,140.82)(497.6,135.35) .. controls (497.65,131.81) and (494.63,128.58) .. (489.84,127.04)(508.15,118.17) .. controls (507.37,119.38) and (506.19,120.44) .. (504.69,121.29)(499.79,106.66) .. controls (499.92,107.15) and (499.98,107.64) .. (499.97,108.14)(479.52,103.06) .. controls (478.79,103.63) and (478.2,104.26) .. (477.75,104.94)(461.9,104.17) .. controls (461.51,104.68) and (461.22,105.22) .. (461.04,105.79)(441.71,106.23) .. controls (442.84,106.67) and (443.88,107.19) .. (444.81,107.8)(417.57,116.9) .. controls (417.68,117.46) and (417.87,118.02) .. (418.11,118.56) ;
%Straight Lines [id:da6692566829495393] 
\draw [line width=2.25]    (460.17,69.5) -- (460.17,95.33) ;
\draw [shift={(460.17,100.33)}, rotate = 270] [fill={rgb, 255:red, 0; green, 0; blue, 0 }  ][line width=0.08]  [draw opacity=0] (16.07,-7.72) -- (0,0) -- (16.07,7.72) -- (10.67,0) -- cycle    ;
%Straight Lines [id:da1598231540463657] 
\draw [line width=2.25]    (366.88,70) -- (461.63,70) ;

%Straight Lines [id:da49374983479675427] 
\draw [line width=2.25]    (460.83,179.67) -- (374.88,179.67) ;
\draw [shift={(369.88,179.67)}, rotate = 360] [fill={rgb, 255:red, 0; green, 0; blue, 0 }  ][line width=0.08]  [draw opacity=0] (16.07,-7.72) -- (0,0) -- (16.07,7.72) -- (10.67,0) -- cycle    ;
%Straight Lines [id:da46117849081262563] 
\draw [line width=2.25]    (460.83,152.33) -- (460.83,179.67) ;
%Straight Lines [id:da469750703967021] 
\draw [line width=2.25]    (460.83,179.67) -- (460.83,241.5) ;
%Straight Lines [id:da5841375949526512] 
\draw [line width=2.25]    (462.38,241) -- (374.88,241) ;
\draw [shift={(369.88,241)}, rotate = 360] [fill={rgb, 255:red, 0; green, 0; blue, 0 }  ][line width=0.08]  [draw opacity=0] (16.07,-7.72) -- (0,0) -- (16.07,7.72) -- (10.67,0) -- cycle    ;

%Straight Lines [id:da8547650862210516] 
\draw [line width=2.25]  [dash pattern={on 6.75pt off 4.5pt}]  (367.38,124.5) -- (402.88,124.5) ;
\draw [shift={(407.88,124.5)}, rotate = 180] [fill={rgb, 255:red, 0; green, 0; blue, 0 }  ][line width=0.08]  [draw opacity=0] (10.36,-4.98) -- (0,0) -- (10.36,4.98) -- (6.88,0) -- cycle    ;
%Flowchart: Alternative Process [id:dp3658455928452795] 
\draw  [dash pattern={on 6.75pt off 4.5pt}][line width=2.25]  (295.4,112.4) .. controls (295.4,108.53) and (298.53,105.4) .. (302.4,105.4) -- (358.4,105.4) .. controls (362.27,105.4) and (365.4,108.53) .. (365.4,112.4) -- (365.4,138.4) .. controls (365.4,142.27) and (362.27,145.4) .. (358.4,145.4) -- (302.4,145.4) .. controls (298.53,145.4) and (295.4,142.27) .. (295.4,138.4) -- cycle ;
%Straight Lines [id:da9139736981425436] 
\draw [line width=2.25]    (200.99,180.86) -- (200.99,103.5) ;
%Straight Lines [id:da3479280318091883] 
\draw [line width=2.25]    (294.28,180.31) -- (199.53,180.36) ;
%Straight Lines [id:da0862182884695365] 
\draw [line width=2.25]    (200.67,240.42) -- (286.63,240.42) ;
\draw [shift={(291.63,240.42)}, rotate = 180] [fill={rgb, 255:red, 0; green, 0; blue, 0 }  ][line width=0.08]  [draw opacity=0] (16.07,-7.72) -- (0,0) -- (16.07,7.72) -- (10.67,0) -- cycle    ;
%Straight Lines [id:da8350092475396569] 
\draw [line width=2.25]    (201.03,241.82) -- (201.03,179.4) ;
%Straight Lines [id:da04073816016384446] 
\draw [line width=2.25]    (199.53,70) -- (286.78,70) ;
\draw [shift={(291.78,70)}, rotate = 180] [fill={rgb, 255:red, 0; green, 0; blue, 0 }  ][line width=0.08]  [draw opacity=0] (16.07,-7.72) -- (0,0) -- (16.07,7.72) -- (10.67,0) -- cycle    ;
%Straight Lines [id:da9960188706465491] 
\draw [line width=2.25]    (200.99,70) -- (201,106.2) ;

% Text Node
\draw (369.13,95.65) node [anchor=north west][inner sep=0.75pt]  [font=\large]  {$\delta _{y}( t)$};
% Text Node
\draw (162.11,168.1) node [anchor=north west][inner sep=0.75pt]  [font=\large]  {$u( t)$};
% Text Node
\draw (465.13,197.9) node [anchor=north west][inner sep=0.75pt]  [font=\large]  {$\tilde{y}( t)$};
% Text Node
\draw (463.98,48.15) node [anchor=north west][inner sep=0.75pt]  [font=\large]  {$y( t)$};
% Text Node
\draw (420.5,117.42) node [anchor=north west][inner sep=0.75pt]  [font=\large] [align=left] {Network};
% Text Node
\draw (321.6,64.4) node [anchor=north west][inner sep=0.75pt]    {$\mathcal{P}$};
% Text Node
\draw (319.6,119.4) node [anchor=north west][inner sep=0.75pt]    {$\mathcal{A}$};
% Text Node
\draw (321.77,174.9) node [anchor=north west][inner sep=0.75pt]    {$\mathcal{F}$};
% Text Node
\draw (321.6,234.4) node [anchor=north west][inner sep=0.75pt]    {$\mathcal{D}$};

\end{tikzpicture}}
    \caption{System Overview}
    \label{fig:SysOverview}
\end{figure}
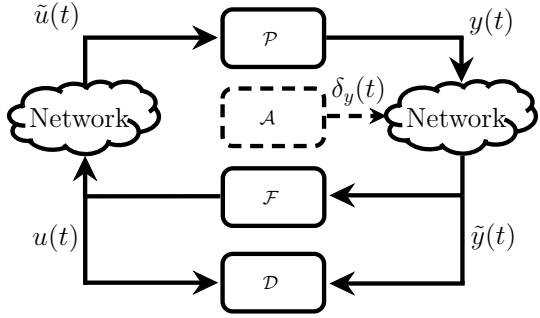

This section introduces the system architecture considered in this paper, as illustrated in Fig.~\ref{fig:SysOverview}, which features an LTI system $\mathcal{P}$ that communicates its output to a local control structure. We consider the scenario in which a malicious Attacker $\mathcal{A}$ may intercept and modify the transmitted sensor data. At the same time, the link between the controller and the actuator is assumed to be secure. This represents a configuration where the controller is implemented locally near the actuator, isolating the control input signals from the vulnerable communication network. Consequently, this work focuses on FDI attacks targeting the sensing layer. The local control structure includes a general output-feedback Detector $\mathcal{D}$ and a dynamic output-Feedback controller $\mathcal{F}$. The feedback detector limits the malicious capabilities of the attacker and characterizes the set of stealthy attacks, while the dynamic output-feedback controller is \textit{designed} to be robust against these stealthy attacks.

%%%%% INCLUDE OTHER SUBSECTIONS %%%%%
\subsection{System Dynamics} \label{sec:SA_SysDynamics}
Consider the LTI perturbed system
\begin{equation} \label{eq:SA_OpenLoop}
    \mathcal{P} \coloneqq \left\{  \begin{aligned}
        \dot{x}(t)  &= A x(t) + B u(t) + G \omega(t),  \\
        y(t)        &= C x(t) + H v(t),  
    \end{aligned} \right.
\end{equation} 
with system state $x(t) \in \mathbb{R}^{n_x}$, control input $u(t) \in \mathbb{R}^{n_u}$, model output $y(t) \in \mathbb{R}^{n_y}$, and unknown system and sensor perturbations $\omega(t) \in \mathbb{R}^{n_\omega}$ and $v(t) \in \mathbb{R}^{n_y}$, respectively, where $n_x, n_u, n_\omega, n_y \in \mathbb{N}$ are the corresponding dimensions. The matrices $A$, $B$, $C$, $G$, and $H$ are of appropriate dimensions, the pair $(A,B)$ is stabilizable, and $(A,C)$ is detectable. The perturbations are assumed to be peak bounded, namely, $\omega(t) \in \mathcal{E}_\omega$ and $v(t)\in \mathcal{E}_v$, with
\begin{subequations} \label{eq:SA_inputSet_original}
    \begin{align}
        \mathcal{E}_\omega  &\coloneqq \{\omega \in \mathbb{R}^{n_\omega}   \mid 1-  \omega^\top W_\omega \omega \geq 0\}, \\
        \mathcal{E}_v &\coloneqq \{v \in \mathbb{R}^{n_v} \mid 1 - v^\top W_v v \geq 0 \},
    \end{align}
\end{subequations}
where $W_\omega \succ 0$ and $W_v \succ 0$ are known.

This manuscript addresses \textit{FDI attacks} in networked control systems. The attacker $\mathcal{A}$ is modeled as a man-in-the-middle adversary that compromises up to $s \in \{1,...,n_y\}$ sensors by injecting malicious data. The corresponding attacker's sensor selection matrix is denoted by $\Gamma \in \{0,1\}^{n_y \times s}$, where each column corresponds to a canonical basis vector indicating a compromised sensor. The corrupted output yields
\begin{align} \label{eq:SA_Sensors}
    \tilde{y}(t) \coloneqq y(t) + \Gamma \delta_y(t),
\end{align}
where $\delta_y(t) \in \mathbb{R}^s$ denotes the additive sensor attack.
\subsection{General Output-Feedback Detector} \label{sec:GOFD}
The system in Fig.~\ref{fig:SysOverview} includes a general output-feedback Detector $\mathcal{D}$ to pinpoint the presence of anomalies. We consider a residual-based detector enabled by a Luenberger state observer, with the estimated state $\hat{x}(t) \in \mathbb{R}^{n_x}$, the residual $r(t) \in \mathbb{R}^{n_y}$, filter gain $L \in \mathbb{R}^{n_x \times n_y}$, and positive definite matrix $\Pi \in \mathbb{R}^{n_y \times n_y}$:
\begin{equation} \label{eq:SA_FeedbackDetector} \mathcal{D} \coloneqq \left\{ \begin{aligned}
    \dot{\hat{x}}(t) &= A \hat{x}(t) + B u(t) + L r(t), \\
    r(t) &= \tilde{y}(t) - C\hat{x}(t), \\
    &\hspace{-6mm} \text{alarm if } r(t)^\top \Pi r(t) > 1.
\end{aligned} \right. \end{equation}
Define the estimation error $e(t) \coloneqq x(t) - \hat{x}(t)$. Given the system dynamics \eqref{eq:SA_OpenLoop} and detector \eqref{eq:SA_FeedbackDetector}, the estimation error dynamics and residual evolve as follows
\begin{equation}\label{eq:SA_ErrorDyn1} \! \! \! \! \! \left\{ \begin{aligned} 
    \dot{e}(t) \! &= \! (A-LC) e(t) \!+\! G \omega(t) \!-\! L H v(t) \!-\! L \Gamma \delta_y(t), \\
    r(t) \!&=\! C e(t) \!+\! H v(t) \!+\! \Gamma \delta_y(t).
\end{aligned} \right. \end{equation}
Because the pair $(A,C)$ is detectable, there exists an $L$ such that $(A-LC)$ is Hurwitz. 

To characterize the residual $r$ under nominal conditions, we introduce the residual set
\begin{equation} \label{eq:SA_residualSet_original} \
    \mathcal{E}_r \coloneqq \{ r(t) \in \mathbb{R}^{n_y} \mid 1- r(t)^\top \Pi r(t) \geq 0 \},
\end{equation}
where $\Pi \succ 0$ defines the shape and size of the residual ellipsoid. This set captures all possible residual trajectories due to the influence of bounded disturbances $\omega(t) \in \mathcal{E}_\omega$ and $v(t) \in \mathcal{E}_v$, as defined in \eqref{eq:SA_inputSet_original}. Whenever the residual leaves $\mathcal{E}_r$, the detector $\mathcal{D}$ raises an alarm, pinpointing the presence of an anomaly. In Appendix~\ref{app:obtianPi}, we provide tools to compute $\Pi$.
\begin{assum}[Existence Residual Set] \label{assum:ExistencePi} \
    We assume the existence of a time instant $t^* \in \mathbb{R}$ at which the FDI $\delta_y(t)$ begins. Prior to this, the system operates under nominal conditions (i.e., $\delta_y(t) = 0, \, \forall t<t^*$) such that the residual $r(t)$ in \eqref{eq:SA_ErrorDyn1} satisfies $r(t) \in \mathcal{E}_r,\,\forall t\geq t^*$, as defined in \eqref{eq:SA_residualSet_original}.  In other words, the transients due to initial conditions have decayed by $t^*$ and $\mathcal{E}_r$ contains all possible residual trajectories due to the bounded disturbances $\omega(t)\in\mathcal{E}_\omega$ and $v(t)\in\mathcal{E}_v$, as defined in \eqref{eq:SA_inputSet_original}.
\end{assum}
\begin{assum}[Stealthy Attack] \label{assum:stealthy_attacker} \
The covert attacker $\mathcal{A}$ is assumed to possess full knowledge of the plant, controller, and detector dynamics. Its objective is to remain undetected, i.e., stealthy, by the anomaly detector $\mathcal{D}$. Accordingly, it selects its attack signal such that the resulting residual signal $r(t)$ in \eqref{eq:SA_ErrorDyn1} is contained inside the monitoring ellipsoidal set $\mathcal{E}_r$, defined in \eqref{eq:SA_residualSet_original}.
\end{assum}

Under Assumption~\ref{assum:stealthy_attacker}, the response of the estimation error dynamics in~\eqref{eq:SA_ErrorDyn1} is not necessarily bounded, as a stealthy attack may destabilize the observer while keeping the residual within the admissible set $\mathcal{E}_r$. Since this paper focuses on improving the robustness of closed-loop stable systems, we restrict attention to systems for which the boundedness of the estimation error under stealthy attacks can be guaranteed. To make this explicit, consider the residual $r(t)$ in~\eqref{eq:SA_ErrorDyn1}. Since $\Gamma$ is full column rank by construction, the attack signal $\delta_y(t)$ can be expressed using the pseudoinverse $\Gamma^\dagger$ as
\begin{align} \label{eq:SD_attackSignal}
    \delta_y(t) = \Gamma^\dagger(r(t) - Ce(t) - H v(t)).
\end{align}
Substituting~\eqref{eq:SD_attackSignal} into~\eqref{eq:SA_ErrorDyn1} yields
\begin{multline} \label{eq:SA_ErrorDyn_residue}
    \dot{e}(t) = \underbrace{\left( A-L(I-\Gamma  \Gamma^\dagger)C\right)}_{A_e} e(t) + G \omega(t) \\
    -L (I - \Gamma \Gamma^\dagger) H v(t) - L \Gamma \Gamma^\dagger r(t),
\end{multline}
where $I - \Gamma \Gamma^\dagger$ denotes the orthogonal projector onto the subspace of measurements not affected by the attack.
%%%
\begin{assum}[Stable Detector Dynamics] \label{assum:ExistenceLforGamma} \
The matrix $\Gamma$ is such that the pair $(A,(I-\Gamma\Gamma^\dagger)C)$ is detectable, i.e.,
\begin{align}
    \operatorname{rank}\!\begin{bmatrix}
    \lambda I - A\\
    (I-\Gamma\Gamma^\dagger)C
    \end{bmatrix} = n_x
    \quad \forall \lambda(A):\ \Re(\lambda)\geq 0.
\end{align}
This condition ensures that the system remains detectable when only the attack-free outputs, characterized by $(I-\Gamma\Gamma^\dagger)C$, are available. Consequently, there exists an observer gain $L$ such that the matrix $A_e$ in~\eqref{eq:SA_ErrorDyn_residue} is Hurwitz. Hence, we can guarantee the boundedness of the estimation error under stealthy attacks.
\end{assum}
\begin{rem}[Choice of Selection Matrix $\Gamma$] \
The matrix $\Gamma$ encodes the attacker's capability by specifying which sensor measurements can be manipulated. When prior knowledge about vulnerable sensors is available, $\Gamma$ can be chosen to represent only those sensors that are considered at risk. In contrast, selecting $\Gamma = I$ corresponds to the conservative assumption that all sensor measurements may be compromised. The choice of $\Gamma$ may be guided by a security allocation strategy (see, e.g.,~\citep{murguia_security_2020}). Regardless of the selection, $\Gamma$ must satisfy Assumption~\ref{assum:ExistenceLforGamma} to ensure that the resulting detector dynamics remain stable under stealthy attacks.
\end{rem}

\subsection{Dynamic Output-Feedback Controller} \label{sec:SA_DOFC}
Consider the dynamic output-Feedback controller $\mathcal{F}$
\begin{equation}\label{eq:SA_DOFC} 
\mathcal{F} \coloneqq
\left\{ \begin{aligned}
    \dot{\rho}(t)   &= A_c \rho(t) + B_c \tilde{y}(t),  \\
    u(t)            &= C_c \rho(t) + D_c \tilde{y}(t),
\end{aligned} \right. \end{equation}
with controller state $\rho(t) \in \mathbb{R}^{n_\rho}$, networked output $\tilde{y}(t) \in \mathbb{R}^{n_y}$, control input $u(t)\in \mathbb{R}^{n_u}$, and controller matrices $A_c$, $B_c$, $C_c$, and $D_c$ of appropriate dimensions. The closed-loop system, \eqref{eq:SA_OpenLoop}, \eqref{eq:SA_Sensors}, \eqref{eq:SA_DOFC} then yields:
\begin{equation} \label{eq:SA_ClosedLoop} \hspace{-2.8mm} \left\{ \begin{aligned} 
    \dot{x}(t) &= (A + B D_c C) x(t) \! + \! B C_c \rho(t) + G \omega(t) \\ 
    & \ \ + B D_c H v(t) +  B D_c \Gamma \delta_y(t), \\ 
    \dot{\rho}(t)  &= B_c C x(t) \! +  A_c \rho(t) \! +  B_c H v(t) \! +  B_c \Gamma \delta_y(t).
\end{aligned} \right. \end{equation}
In this manuscript, we present a framework to optimize a given \emph{base controller} $\mathcal{F}$ (as defined in \eqref{eq:SA_DOFC}) to enhance the robustness of the closed-loop system \eqref{eq:SA_ClosedLoop} against stealthy attacks, while preserving its nominal performance in the attack-free case. Exploiting the system output $y(t)$, we construct a class of equivalent dynamic controller realizations, referred to as \emph{plant equivalent controller (PEC) realizations}, that preserve the nominal input-output behavior of the closed-loop system but differ in their robustness properties. 

\begin{comment}
In this manuscript, we present a framework to optimize a given \emph{base controller} $\mathcal{F}$ (as defined in \eqref{eq:SA_DOFC}) to enhance the robustness of the closed-loop system \eqref{eq:SA_ClosedLoop} against stealthy attacks, while preserving its nominal performance in the attack-free case. Exploiting redundancy in the system output $y(t)$, we construct a class of equivalent dynamic controller realizations, referred to as \emph{plant equivalent controller (PEC) realizations}, that preserve the nominal input-output behavior of the closed-loop system but differ in their robustness properties. 
\end{comment}

To evaluate robustness, we model the anomaly detector $\mathcal{D}$ such that constraints on the attack signal are imposed: to remain stealthy, an attack must satisfy the ellipsoidal detection bounds $r^\top \Pi r<1$. Under these constraints, we consider the set of system trajectories that the attacker can induce. As a security metric, we use the volume of the reachable set of the closed-loop system under this class of stealthy attacks. We then search over all PEC realizations and formulate an optimization problem to minimize the volume of the reachable set. The resulting PEC realization improves robustness against stealthy attacks while preserving the closed-loop performance.

For the remainder of the manuscript, we drop the time dependency notation $(t)$ for national simplicity. Therefore, $x(t) \coloneqq x$, $u(t) \coloneqq u$, $\omega(t) \coloneqq \omega$,  $v(t) \coloneqq v$, $\rho(t) \coloneqq \rho$, $\tilde{y}(t) \coloneqq \tilde{y}$, $\delta_y(t) \coloneqq \delta_y$, and $r(t) \coloneqq r$.

% Version 1 
\begin{comment}
This section introduces the system architecture considered in this paper, as illustrated in Fig.~\ref{fig:SysOverview}, which features an LTI system $\mathcal{P}$ communicating its output to a control structure, where a malicious Attacker $\mathcal{A}$ may intercept and modify the transmitted output data. The control structure includes a general output-feedback Detector $\mathcal{D}$ and a dynamic output-Feedback controller $\mathcal{F}$. The feedback detector limits the malicious capabilities of the attacker and characterizes the set of stealthy attacks, while the dynamic output-feedback controller is \textit{designed} to be robust against these stealthy attacks.
\end{comment}

%% System and controller formulation
\section{Plant Equivalent Controller Realizations} \label{sec:CR}
In this section, we introduce the concept of \emph{plant equivalent controller (PEC) realizations} and examine how different realizations of the base controller impact the robustness of the closed-loop system. 
\begin{assum}[Base Controller] \label{assum:ExistenceDOFC} \hfill %
   Consider the LTI system \eqref{eq:SA_OpenLoop}, where $(A,B)$ is stabilizable and $(A,C)$ is detectable. There exists a dynamic base controller $\mathcal{F}$ of the form \eqref{eq:SA_DOFC} that stabilizes the nominal closed-loop system and achieves the desired behavior in the absence of external disturbances, i.e., $\omega = v = \delta_y = 0$. We refer to $\mathcal{F}$ as the base controller.
\end{assum}
\begin{defn}[Controller Realization] \label{def:Realization} \ 
    Let $\mathcal{F}$ in \eqref{eq:SA_DOFC} be a \emph{stabilizing} base controller with internal state $\rho \in \mathbb{R}^{n_\rho}$, described by the matrices $(A_c, B_c, C_c, D_c)$. A controller $\bar{\mathcal{F}}$ with internal state $\bar{\rho} \in \mathbb{R}^{n_\rho}$ and matrices $(\bar{A}_c, \bar{B}_c, \bar{C}_c, \bar{D}_c)$ is said to be a \emph{realization} of $\mathcal{F}$ if for every initial state $\rho(0)$ there exists an initial state $\bar{\rho}(0)$ such that, for any trajectory $\tilde y$, both controllers generate identical control inputs, i.e., $\bar{u}(t) = u(t)$ for all $t \ge 0$. 
\end{defn}
\begin{rem}[Standard Controller Realization] \label{rem:StandardRealization} \ 
    A standard controller realization can be obtained through a similarity transformation of the controller state. Let $\bar \rho = S \rho$ for some invertible matrix $S$, which yields the controller matrices: $\bar{A}_c = S A_c S^{-1}$, $\bar{B}_c = S B_c$, $\bar{C}_c = C_c S^{-1}$, and $\bar{D}_c = D_c$. Under this similarity transformation, the controller $\bar{\mathcal{F}}$ produces the same control input as the base controller $\mathcal{F}$ for any trajectory of $\tilde y$. 
\end{rem}
A standard controller realization of $\mathcal{F}$ preserves its full input--output behavior for any measurement signal $\tilde y$, and consequently, for any disturbance. In our setting, however, $\mathcal{F}$ is designed for a specific plant, and full input--output equivalence is not required. We only require that a realization preserves the \emph{nominal} closed-loop behavior (i.e., for $\omega = v = \delta_y = 0$). 

Using the plant dynamics and available sensor information, we show that the internal controller coordinates can be transformed to incorporate (part of) the plant state. Such transformations maintain the nominal closed-loop behavior with the given plant, but no longer guarantee identical behavior in the presence of disturbances. The resulting class of realizations we refer to as \emph{PEC realizations}. While all PEC realizations exhibit the same nominal closed-loop behavior, their robustness properties differ for $\omega \neq 0$, $v \neq 0$, or $\delta_y \neq 0$, which is a property we exploit in the remainder of the paper.

To derive the PEC realizations, we first characterize the components of the plant state that can be algebraically reconstructed from $y$ in the disturbance-free case, which determines the available degrees of freedom. By applying a linear change of coordinates to the controller state of $\mathcal{F}$, combined with the plant dynamics, we obtain the PEC realizations in Section~\ref{sec:eqControllers}, and derive the corresponding closed-loop dynamics in Section~\ref{sec:eqClosedLoop}.

\subsection{Class of Plant Equivalent Controller Realizations} \label{sec:eqControllers} 
When defining a class of equivalent controller realizations, it is essential that each controller realization can be implemented using the measured plant output. To this end, we distinguish which components of the state $x$ can be reconstructed from the output $y$ algebraically.

Consider the following change of coordinates:
\begin{align} \label{eq:CR_C_xbar}
    \bar{x} = \begin{bmatrix}
        \bar{x}_1 \\ \bar{x}_2
    \end{bmatrix} \coloneqq \begin{bmatrix}
        T_1 \\ T_2
    \end{bmatrix} x = T x,
\end{align}
where $T_1 \in \mathbb{R}^{n_y \times n_x}$ is chosen such that its rows form a basis of the row space of $C$, with $C$ as defined in \eqref{eq:SA_OpenLoop}. The matrix $T_2 \in \mathbb{R}^{(n_x-n_y) \times n_x}$ is chosen such that its rows form a basis of $\ker(C)$, ensuring that the transformation matrix $T$ is of full rank. Here $\bar{x}_1$ represents the state component that can be reconstructed (algebraically) from the output $y$, while the remaining state component, $\bar{x}_2$, corresponds to directions in $\ker{(C)}$ and therefore cannot be reconstructed (algebraically) from the output. 
\begin{assum}[Full Row Rank of $C$] \
    Without loss of generality, we assume that the system matrix $C$ in~\eqref{eq:SA_OpenLoop} has full row rank, which allows us to simplify the change of coordinates to $T = [C^\top \; T_2^\top]^\top$. If $C$ is rank-deficient, it can always be reduced to a full row rank matrix by removing linearly dependent rows, which does not affect its row space and therefore does not affect the analysis.
\end{assum}

The system dynamics in the new coordinates $\bar{x}$ are obtained by applying the coordinate transformation in \eqref{eq:CR_C_xbar} to \eqref{eq:SA_OpenLoop} for $T = [C^\top \; T_2^\top]^\top$:
\begin{subequations}\label{eq:CR_C_dxbar_OL} \begin{align} 
    \dot{\bar{x}} &= T A T^{-1} \bar{x} + T B u + T G \omega  
    = \bar{A} \bar{x} + \bar{B} u + \bar{G} \omega, \\
        y &= CT^{-1}\bar{x} + Hv = \bar{C} \bar{x} + Hv,  
\end{align}
which gives the structure:
 \begin{align} 
   \begin{bmatrix}
       \dot{\bar{x}}_1 \\ \dot{\bar{x}}_2
   \end{bmatrix} &=
    \begin{bmatrix}
        \bar{A}_{11} & \bar{A}_{12} \\ 
        \bar{A}_{21} & \bar{A}_{22}
    \end{bmatrix} 
    \begin{bmatrix}
         \bar{x}_1 \\ \bar{x}_2
    \end{bmatrix}
    + 
    \begin{bmatrix}
        \bar{B}_1 \\  \bar{B}_2
    \end{bmatrix} u
        + 
    \begin{bmatrix}
        \bar{G}_1 \\  \bar{G}_2
    \end{bmatrix} \omega, \\
    y &= \begin{bmatrix}
        I & 0
    \end{bmatrix} \bar{x} + Hv.
\end{align} \end{subequations}
The change of coordinates proposed in \eqref{eq:CR_C_xbar} isolates the part of the state that can be reconstructed from the output and subsequently used in realizing the base controller $\mathcal{F}$. Specifically, $\bar{x}_1$ and the dynamics in \eqref{eq:CR_C_dxbar_OL} form the basis for deriving the PEC realizations. 
\begin{prop}[PEC Realizations] \label{prop:controllerRealizations} \
    Consider the base controller~\eqref{eq:SA_DOFC}, the LTI system dynamics~\eqref{eq:CR_C_dxbar_OL}, new controller state $\bar{\rho} \in \mathbb{R}^{n_\rho}$, plant equivalent controller (PEC)
    realization matrix $F \in \mathbb{R}^{n_\rho \times n_y}$, and the change of coordinates in controller state $\rho$
    \begin{align} \label{eq:CR_Transformation}
        \bar{\rho} &= \rho + F \bar{x}_1.
    \end{align}
    Then, this change of coordinates yields a PEC realization, i.e., it preserves the control input of~\eqref{eq:SA_DOFC} and, hence, it also preserves the closed-loop behavior in the absence of external disturbances ($\omega = v = \delta_y = 0$), if and only if $F$ satisfies $F \bar{A}_{12} = 0$. For any such $F$, the resulting PEC realization is given by:
    \begin{subequations} \label{eq:CR_C_Fbar}
    \begin{align}
    \begin{split}
       \dot{\bar{\rho}} &= (A_c + F \bar{B}_1 C_c) \bar{\rho} + \left(B_c - A_c F\right. \\ 
        & \qquad \left.  + F \bar{A}_{11} + F \bar{B}_1 D_{c} - F \bar{B}_1 C_c F \right) y, 
    \end{split}\\
        u &= C_c \bar{\rho} + (D_{c} - C_c F) y. 
    \end{align} \end{subequations}
    \end{prop}
\begin{pf}
    Consider the base controller~\eqref{eq:SA_DOFC} and the LTI system dynamics in transformed coordinates~\eqref{eq:CR_C_dxbar_OL}. The transformed state satisfies $\bar{x}_1 = Cx =y$, and $\bar{x}_2 = T_2 x$, with $C$ as defined in~\eqref{eq:SA_OpenLoop} and $T_2 \subseteq \ker(C)$. Hence, $\bar{x}_1$ and $\bar{x}_2$ correspond to components of the state associated with orthogonal subspaces in the original state space.
    
    For $\omega = v = \delta_y = 0$, it follows from \eqref{eq:SA_DOFC},~\eqref{eq:CR_C_dxbar_OL}, and \eqref{eq:CR_Transformation}:
    \begin{subequations} \begin{align} \label{eq:CR_C_Fbar_u}
        u &= C_c \bar{\rho} + (D_{c} - C_c F) \bar{x}_1 \\
        \intertext{with the dynamics given by $\dot{\bar{\rho}} = \dot{\rho} + F\dot{\bar{x}}_1$:}  
    \begin{split} \label{eq:CR_C_Fbar_rho1}
            \dot{\bar{\rho}} &= A_c\left(\bar{\rho} - F \bar{x}_1\right) + B_c \bar{x}_1 \\
            & \quad +F\left(\bar{A}_{11} \bar{x}_1 + \bar{A}_{12} \bar{x}_2 + \bar{B}_1 u \right).
        \end{split}
    \end{align}\end{subequations}
     These dynamics still depend on $\bar{x}_2$, whereas only $\bar{x}_1$ is measurable. Therefore, to ensure that the PEC realization depends solely on the system output $y$, it is required that $F \bar A_{12} = 0$, i.e., columns of $F$ must lie in the left null space of $\bar{A}_{12}$. Substituting~\eqref{eq:CR_C_Fbar_u} into~\eqref{eq:CR_C_Fbar_rho1} yields the PEC realizations in~\eqref{eq:CR_C_Fbar}, where $\bar{x}_1 \coloneqq y$. \hfill $\blacksquare$ 
\end{pf}
\begin{rem}[Proposed Transformation] \
    The change of coordinates~\eqref{eq:CR_Transformation} may be combined with any similarity transformation of the controller state as in Remark~\ref{rem:StandardRealization}. In particular, one may apply an invertible matrix $S$ to obtain $\bar{\rho} = S \rho + SF \bar{x}_1$. Without loss of generality, we set $S = I$ in Proposition~\ref{prop:controllerRealizations}, since this does not affect the resulting behavior of the PEC realization.
\end{rem}
Proposition~\ref{prop:controllerRealizations} imposes the constraint $F \bar{A}_{12} = 0$. The conditions under which such a matrix $F$ exists are established in the following lemma.
\begin{lem}[Existence PEC Realization] \label{lem:ExistsenceF} \
    Consider the LTI system \eqref{eq:CR_C_dxbar_OL}, the proposed change of coordinates \eqref{eq:CR_Transformation}, and the PEC realizations \eqref{eq:CR_C_Fbar}. 
    Let $\bar A_{12} \in \mathbb{R}^{n_y \times (n_x-n_y)}$ and define 
    \begin{align}
        p \coloneqq \dim\big(\ker(\bar A_{12}^\top)\big) = n_y - \rank(\bar A_{12}).
    \end{align}
    Then, there exists a nontrivial matrix $F\in\mathbb{R}^{n_\rho\times n_y}$ satisfying $F\bar A_{12} = 0$ if and only if $p>0$.
\end{lem}
\begin{pf}
    By the rank-nullity theorem~\citep{leon_linear_2021}, the left null space of $\bar A_{12}$ has dimension
    \begin{align}
        p \coloneqq \dim\left(\ker(\bar A_{12}^\top)\right) = n_y - \rank(\bar A_{12}^\top).
    \end{align}
    Hence, a nontrivial $F$ satisfying $F \bar A_{12} = 0$ exists if and only if $p > 0$. In that case, the rows of $F^\top$ can be chosen from a basis of $\ker(\bar A_{12}^\top)$, i.e., $F^\top \subseteq \ker(\bar A_{12}^\top)$, which implies $F \bar A_{12} = 0$. \null\hfill $\blacksquare$
\end{pf}

\begin{rem}[Dynamic Redundancy] \
    The existence of a nontrivial matrix $F$ satisfying $F\bar{A}_{12}=0$ in Lemma~\ref{lem:ExistsenceF} does not require hardware redundancy (i.e., multiple sensors measuring the same state). Instead, it arises from the dynamic coupling present in the system output, i.e., different sensors measure state variables that are related through their dynamics. This allows the PEC realization to combine measurements of distinct but dynamically coupled state variables, such as, e.g., position, velocity, and acceleration in mechanical systems. Consequently, PEC realizations leverage the underlying dynamics (physics) of the system instead of requiring multiple sensors measuring the same state. 
\end{rem}  
\begin{rem}[Lemma~\ref{lem:ExistsenceF} as Design Tool] \    
    In practice, the condition ($p>0$) of Lemma~\ref{lem:ExistsenceF} is often satisfied in systems with sufficient sensors, as is common in, e.g., automotive or aerospace applications. If this condition does not hold for a given set of sensors, then the base controller $\mathcal{F}$ is the only admissible PEC realization. In this case, Lemma~\ref{lem:ExistsenceF} can serve as a design tool, indicating which additional sensor(s) would enable the use of PEC realizations.
\end{rem}

\subsection{Equivalent Closed-Loop System Formulation}\label{sec:eqClosedLoop}

Consider the closed-loop state vector $\zeta \coloneqq \begin{bmatrix} \bar{x}_1^\top & \bar{x}_2^\top & \rho^\top \end{bmatrix}^\top$, where $\zeta \in \mathbb{R}^{n_\zeta}$ and $n_\zeta = n_x + n_\rho$. In the absence of external disturbances, i.e., $\omega = v = \delta_y = 0$, the closed-loop dynamics from~\eqref{eq:SA_DOFC} and~\eqref{eq:CR_C_dxbar_OL} yield
    \begin{align} \label{eq:CR_C_dzeta}
        \dot{\zeta} &= \mathcal{A} \zeta, \quad
    \mathcal{A} = \begin{bmatrix}
        \bar{A}_{11} + \bar{B}_1 D_c    & \bar{A}_{12}  & \bar{B}_1 C_c \\ 
        \bar{A}_{21} + \bar{B}_2 D_c    & \bar{A}_{22}  & \bar{B}_2 C_c \\
        B_c                             & 0             & A_c
    \end{bmatrix},
\end{align}
which, under Assumption~\ref{assum:ExistenceDOFC}, guarantees the desired nominal performance.
\begin{assum}[Existence PEC Realizations] \label{assum:ExistenceF} \
   For~\eqref{eq:CR_C_dzeta}, it is assumed that $p =\dim\left(\ker(\bar A_{12}^\top)\right)>0$.
\end{assum}
Under Assumption~\ref{assum:ExistenceF}, there exists a PEC realization that yields an equivalent closed-loop system to~\eqref{eq:CR_C_dzeta} when $\omega=v=\delta_y=0$, see Lemma~\ref{lem:ExistsenceF}. This equivalence, however, only pertains to the nominal behavior. Namely, the PEC realization replaces the nominal controller with a controller that utilizes a different sensor configuration, and therefore, the disturbances enter the closed-loop dynamics differently. To make this explicit, we first substitute $\tilde{y}$ for $y$ in~\eqref{eq:CR_C_Fbar} and obtain the PEC realization including disturbances:
 \begin{equation} \label{eq:CR_C_Fbar_att}
\! \! \! \bar{\mathcal{F}} \! \coloneqq \!
\left\{ \! 
\begin{aligned}
 \dot{\bar{\rho}} &= (A_c + F \bar{B}_1 C_c) \bar{\rho} + \left( B_c - A_c F \right.\\ 
    &  \left.+ F \bar{A}_{11} + F \bar{B}_1 D_{c} - F \bar{B}_1 C_c F \right) \tilde{y}, \\
    u &= C_c \bar{\rho} + (D_{c} - C_c F) \tilde{y},
\end{aligned} \right. \end{equation}
where $\tilde{y} = \bar{x}_1 + H v + \Gamma \delta_y$. 

Next, we aim to obtain the closed-loop dynamics in the coordinates $\zeta$, hence the original controller state, but using the PEC realization in~\eqref{eq:CR_C_Fbar_att}. To do so, \eqref{eq:CR_C_Fbar_att} is first expressed in terms of the original controller state $\rho$ by applying the inverse transformation of~\eqref{eq:CR_Transformation}, resulting in
\begin{subequations}\label{eq:CR_C_Fbar_PEC}\begin{align}\begin{split}
    u &= C_c \rho + D_{c} \bar{x}_1 + \\
      & \qquad + (D_{c} - C_c F)H v +(D_{c} - C_c F)\Gamma  \delta_y.
\end{split}
\intertext{Furthermore, using $\dot{\rho} = \dot{\bar{\rho}} - F \dot{\bar{x}}_1$, and substituting $\dot{\bar{x}}_1$ from~\eqref{eq:CR_C_dxbar_OL} with $\omega \neq 0$, we obtain}
\begin{split} \label{eq:CR_drhoInverse}
    \dot{\rho} &= (A_c + F \bar{B}_1 C_c) (\rho + F \bar{x}_1) + \left( B_c - A_c F \right. \\ 
    & \quad \left.+ F \bar{A}_{11} + F \bar{B}_1 D_{c} - F \bar{B}_1 C_c F \right) (\bar{x}_1 + Hv  \\
    & \quad + \Gamma \delta_y) - F\left(\bar{A}_{11} \bar{x}_1 + \bar{A}_{12} \bar{x}_2 + \bar{B}_1 u + \bar{G}_1 \omega \right) \\
    &= A_c \rho + B_c \bar{x}_1 - F \bar{G}_1 \omega ,\\
    & \quad + (B_c - A_c F + F \bar{A}_{11})(Hv + \Gamma \delta_y).
\end{split}\end{align}\end{subequations}
Substituting~\eqref{eq:CR_C_Fbar_PEC} into~\eqref{eq:CR_C_dxbar_OL} yields the closed-loop system
\begin{subequations} \label{eq:CR_C_dzeta_att}
    \begin{align} 
        \dot{\zeta} &= \mathcal{A} \zeta + \mathcal{G} \omega + \mathcal{H} v + \mathcal{T} \delta_y
    \end{align}
with $\mathcal{A}$ as defined in \eqref{eq:CR_C_dzeta} and
    \begin{align}
    \mathcal{G} &= 
    \begin{bmatrix}
        \bar{G}_1\\ 
        \bar{G}_2\\
        -F\bar{G}_1
    \end{bmatrix},    
        \mathcal{H} = 
    \begin{bmatrix}
        \bar{B}_1(D_c - C_c F) H\\ 
        \bar{B}_2(D_c - C_c F) H\\
        (B_c - A_c F + F \bar{A}_{11}) H
    \end{bmatrix},\\
        \mathcal{T} &= 
    \begin{bmatrix}
        \bar{B}_1(D_c - C_c F) \Gamma \\ 
        \bar{B}_2(D_c - C_c F) \Gamma \\
        (B_c - A_c F + F \bar{A}_{11})\Gamma
    \end{bmatrix} \label{eq:CR_C_dzeta_att_T}.
    \end{align}
\end{subequations}
\begin{rem}[Equivalent Closed-Loop System] \
    	For any PEC realization matrix $F$, the autonomous part of the closed-loop dynamics in~\eqref{eq:CR_C_dzeta_att} is governed by the same matrix $\mathcal{A}$ as in~\eqref{eq:CR_C_dzeta}. The choice of $F$ only modifies how the external disturbances $(\omega, v, \delta_y)$ enter the dynamics. Consequently, in the absence of disturbances, the closed-loop behavior is preserved and invariant under the transformation~\eqref{eq:CR_Transformation}.
\end{rem}
The system in~\eqref{eq:CR_C_dzeta_att} explicitly captures the influence of the process disturbance $\omega$, sensor noise $v$, and FDI attacks $\delta_y$. Moreover, the matrices ${\mathcal{G}}$, ${\mathcal{H}}$, and ${\mathcal{T}}$, which map the external disturbances to the closed-loop state evolution, depend linearly on the PEC realization matrix $F$, which enables the use of linear and convex optimization methods for its design. 

Summarizing,~\eqref{eq:CR_C_dzeta_att} extends the nominal closed-loop system in~\eqref{eq:CR_C_dzeta} by incorporating the effects of process and measurement disturbances, as well as FDI attacks. The robustness against these disturbances now explicitly depends on the chosen PEC realization. Here, the closed-loop system is driven by the FDI attack; however, the adversary strategy is unknown beyond the assumption of stealthiness (Assumption~\ref{assum:stealthy_attacker}). In the following section, we include a detector $\mathcal{D}$ and reformulate~\eqref{eq:CR_C_dzeta_att} to be driven by the residual $r$ rather than the FDI attack signal $\delta_y$.

\section{Optimal Plant Equivalent Controller Realization against Stealthy Attacks} \label{sec:DI}
In this section, we provide tools to quantify (for a given $\mathcal{P},\, \mathcal{D}$, and $\mathcal{F}$) and minimize (by optimizing for $F$) the impact of the attack $\delta_y$ on the system when the anomaly detector \eqref{eq:SA_FeedbackDetector} is used for attack detection. Moreover, we focus on the class of attacks that keep the monitor from raising alarms, i.e., stealthy attacks (see Assumption~\ref{assum:stealthy_attacker}). Here, we characterize ellipsoidal bounds on the states that stealthy attacks can induce in the system. In particular, we provide tools based on Linear Matrix Inequalities (LMIs) to compute ellipsoidal bounds on the reachable set of the attack sequence given the system dynamics, the control strategy, the detector, and the set of sensors being attacked. Afterwards, these tools are exploited to find $F$ such that the ellipsoidal bound is minimized, reducing the impact $\delta_y$ has on the closed-loop system.   

To prepare for this analysis, we rewrite the detector~\eqref{eq:SA_ErrorDyn1} in the closed-loop coordinates of~\eqref{eq:CR_C_dzeta_att}. Using the coordinate transformation in~\eqref{eq:CR_C_xbar} with $T = [C^\top, T_2^\top]^\top$, the detector's error dynamics become
\begin{equation} \label{eq:DI_FeedbackDetector_bar} \left\{ \begin{aligned}
    \dot{\bar{e}} &= \left(\bar A- \begin{bmatrix} \bar{L} & 0 \end{bmatrix} \right) \bar{e} + \bar{G} \omega - \bar{L} H v  - \bar{L} \Gamma \delta_y, \\
    r &= \bar e_1 + H v + \Gamma \delta_y,
\end{aligned} \right. \end{equation}
with $\bar{e} \! = \! [\bar e_1^\top \, \bar e_2^\top]^\top \!\! \coloneqq \! [(\bar x_1 - \hat{\bar{x}}_1)^\top \, (\bar x_2 - \hat{\bar{x}}_2)^\top]^\top\!\!,\,$ and $\bar{L} \! \coloneqq \! TL$.

Since $\Gamma$ in~\eqref{eq:DI_FeedbackDetector_bar} is full column rank by construction, and using~\eqref{eq:DI_FeedbackDetector_bar}, the attack signal $\delta_y$ can be expressed via the pseudoinverse $\Gamma^\dagger$ as 
\begin{align} \label{eq:DI_attackSignal}
    \delta_y = \Gamma^\dagger(r - \bar{e}_1 - H v). 
\end{align}
Substituting~\eqref{eq:DI_attackSignal} into the closed-loop dynamics~\eqref{eq:CR_C_dzeta_att}:
 \begin{align} \label{eq:DI_ClosedLoop_residue}
    \dot{\zeta} = \mathcal{A} \zeta + \mathcal{G} \omega + (\mathcal{H} - \mathcal{T} \Gamma^\dagger H) v + \mathcal{T} \Gamma^\dagger r - \mathcal{T} \Gamma^\dagger \bar{e}_1,
\end{align}
while the error dynamics in~\eqref{eq:DI_FeedbackDetector_bar} become
\begin{multline} \label{eq:DI_ErrorDyn_residue}
    \dot{\bar{e}} = \underbrace{\left(\bar A- \begin{bmatrix} \bar{L}(I-\Gamma  \Gamma^\dagger) & 0 \end{bmatrix} \right)}_{\bar{A}_e} \bar{e} + \bar{G} \omega \\ -\bar{L} (I - \Gamma \Gamma^\dagger) H v - \bar{L} \Gamma \Gamma^\dagger r. 
\end{multline}
The signals $\omega \in \mathcal{E}_\omega$, $v \in \mathcal{E}_v$, and $r \in \mathcal{E}_r$ are bounded and belong to known sets \eqref{eq:SA_inputSet_original} and \eqref{eq:SA_residualSet_original}, respectively, under the stealthy attack assumption, Assumption~\ref{assum:stealthy_attacker}.

A difficulty is that the closed-loop dynamics in~\eqref{eq:DI_ClosedLoop_residue} are coupled to the detector through $\bar{e}_1$, so the resulting combined system matrix depends on $\mathcal{T}$, and hence on $F$, see~\eqref{eq:CR_C_dzeta_att_T}. Optimizing $F$ directly to minimize the reachable set of the coupled system would therefore yield a nonlinear and nonconvex problem. To circumvent this problem, this section proposes a solution in two stages. In Section~~\ref{sec:ADP}, we analyze~\eqref{eq:DI_ErrorDyn_residue} and characterize its reachable set via the ellipsoidal set $\bar{\mathcal{E}}_e \coloneqq \{ \bar e \in \mathbb{R}^{n_x} \mid 1- \bar{e}^\top \bar P_e \bar e \geq 0\}$, which is independent of the PEC realization matrix $F$. Subsequently, in Section~\ref{sec:CRIDS}, the influence of $\bar e_1 \in \bar{\mathcal{E}}_e$ on~\eqref{eq:DI_ClosedLoop_residue} is analyzed, leading to an LMI that minimizes the ellipsoidal reachable set of $\zeta$ and yields an \emph{optimal PEC realization}, characterized by $F$.

\subsection{Anomaly Detector Performance}\label{sec:ADP}\vspace{-2mm}
The objective here is to quantify the reachable set $\bar{\mathcal{R}}_e (t)$ (Definition~\ref{defn:reachableSet}), which contains all estimation error trajectories $\bar e(t)$ that originate from the initial condition $\bar e(0) = \bar e_0\in \mathbb{R}^{n_x}$, and evolve under all admissible perturbations $\omega \in \mathcal{E}_\omega$, $v \in \mathcal{E}_v$, and $r \in \mathcal{E}_r$, i.e., 
\begin{align} \label{eq:DI_RS_eBar}
    \bar{\mathcal{R}}_e (t) := \left\{ \bar e (t) \, \middle| \,
    \begin{aligned}
        & \exists\, \omega(s)\in \mathcal{E}_\omega, \, v(s)\in \mathcal{E}_v,\\ 
        & r(s)\in \mathcal{E}_r \text{ s.t. } \bar e(s) \\
        &  \text{ solution to } \eqref{eq:DI_ErrorDyn_residue}, \, s \in [0,t]
    \end{aligned}
    \right\}.
\end{align}
Computing $\bar{\mathcal{R}}_e (t)$ exactly is generally not tractable. However, since the signals $\omega$, $v$, and $r$ are bounded, the infinite-time reachable set $\bar{\mathcal{R}}_e^\infty \coloneqq \lim_{t \to \infty} \bar{\mathcal{R}}_e (t)$ admits an outer ellipsoidal approximation. Specifically, there exists a matrix $\bar P_e \succ 0$, such that $\! \bar{\mathcal{R}}_e^\infty  \subseteq  \bar{\mathcal{E}}_e  \coloneqq  \{  \bar e \in  \mathbb{R}^{n_x} \mid  1 - \bar e^\top \bar P_e \bar e \geq 0 \}$. Under Assumption~\ref{assum:ExistenceLforGamma}, the infinite-time reachable set $\bar{\mathcal{R}}_e^\infty$ is well-defined. 

Furthermore, since the closed-loop dynamics in~\eqref{eq:DI_ClosedLoop_residue} depend only on the component $\bar e_1$ of the estimation error, the performance of the system is fully determined by this subspace. Therefore, we focus on minimizing the projection of $\bar{\mathcal{E}}_e$ onto the $\bar e_1$ subspace in the following lemma.
\begin{lem}[Detector's Ellipsoidal Set] \label{lemma:ErrorSet} \ 
    Let Assumption~\ref{assum:ExistenceLforGamma} hold, consider the perturbed error dynamics~\eqref{eq:DI_ErrorDyn_residue} and the reachable set~\eqref{eq:DI_RS_eBar}. For a given constant $\bar \alpha_e\in \mathbb{R}_{\geq0}$, if there exists matrices $\bar P_e \in \mathbb{R}^{n_x \times n_x}$, $Z \in \mathbb{R}^{n_y \times n_y}$, and constants $\bar \beta_{e,\omega}, \bar\beta_{e,v}, \bar\beta_{e,r} \in \mathbb{R}_{\geq0}$, being the solution to the convex program:
         \begin{subequations} \label{eq:lemma_ErrorSet}
            \begin{equation} \label{eq:lemma_ErrorSet1}
                \left\{\!\begin{aligned}
                    &\min_{Z, \bar{P}_{e},\bar{\beta}_{e,\omega}, \bar{\beta}_{e,v}, \bar{\beta}_{e,r}} \trace(Z),\\
            	    &\text{\emph{s.t.}}  \quad  \bar{\beta}_{e,\omega}, \bar{\beta}_{e,v}, \bar{\beta}_{e,r}\in\mathbb{R}_{\geq0}, \\
                    &\begin{bmatrix} Z & \bar{C}_1 \\ \bar{C}_1^\top & \bar{P}_e \end{bmatrix} \succeq 0, \quad \bar P_{e}\succ 0,\quad Z \succ 0,\\
                    &-\bar{\mathcal M}_{e} - \bar{\alpha}_{e} \bar{\mathcal N}_{e} - \bar{\beta}_{e,\omega} \mathcal{S}_{\omega, \bar\kappa} \\ 
                    & \hspace{10mm} - \bar{\beta}_{e,v} \mathcal{S}_{v, \bar\kappa} - \bar{\beta}_{e,r} \mathcal{S}_{r,\bar\kappa}  \succeq 0,
                \end{aligned}\right.
            \end{equation}
    with matrices 
            \begin{align}
                \bar{C}_1 &\coloneqq \begin{bmatrix}
                    I_{n_y} & 0
                \end{bmatrix}, \quad
                 \bar{ \mathcal N}_e \coloneqq  \diag \begin{bmatrix} \bar P_e, 0, 0, 0, -1 \end{bmatrix}, \\
                 \bar{\mathcal M}_e &\coloneqq  \begin{bmatrix}
                    \bar A_e^\top \bar P_e + \bar P_e \bar A_e & * & *  & * & * \\
                    \bar G^\top \bar{P}_e & 0 & * & * & * \\
                    -(\bar L(I-\Gamma \Gamma^\dagger) H)^\top \bar{P}_e & 0 & 0 & * & * \\
                    - (\bar L \Gamma \Gamma^\dagger)^\top \bar{P}_e & 0 & 0 & 0 & * \\
                    0 & 0 & 0 & 0 & 0 \\
                \end{bmatrix}, \\
                \mathcal{S}_{\omega,\bar \kappa} &\coloneqq  \diag \begin{bmatrix} 0, -W_\omega, 0, 0, 1 \end{bmatrix},  \\
                \mathcal{S}_{v,\bar \kappa} &\coloneqq  \diag \begin{bmatrix} 0, 0, -W_v, 0, 1 \end{bmatrix}, \\
                \mathcal{S}_{\bar \kappa, r} &\coloneqq  \diag \begin{bmatrix} 0, 0, 0, -\Pi, 1 \end{bmatrix},
            \end{align}    
        \end{subequations}
    then, the ellipsoidal set 
    \begin{align} \label{eq:CR_errorBarEllipse}
        \bar{\mathcal{E}}_e := \left\{ \bar{e} \in \mathbb{R}^{n_x} \mid 1- \bar{e}^\top \bar{P}_{e} \bar{e} \geq 0 \right\}, 
    \end{align}
    is forward-invariant and contains the infinite-time reachable set $\bar{\mathcal{R}}_e^\infty \coloneqq \lim_{t \to \infty} \bar{\mathcal{R}}_e (t) \subseteq \bar{\mathcal{E}}_e$. The projection of $\bar{\mathcal{E}}_e$ onto the $\bar{e}_1$ subspace is contained in the ellipsoid
    \begin{equation} \label{eq:e1_ellipse}
        \bar{\mathcal{E}}_{e_1} := \left\{ \bar{e}_1 \in \mathbb{R}^{n_y} \mid 1- \bar{e}_1^\top Z^{-1} \bar{e}_1 \geq 0 \right\},
    \end{equation}
    which has the minimal sum of semi-axes among all ellipsoidal outer approximations.
\end{lem}
\begin{pf} 
    The proof can be found in Appendix~\ref{app:proofDetectorEllipsoid1}.
\end{pf}
Lemma~\ref{lemma:ErrorSet} provides an ellipsoidal outer approximation such that $\bar{\mathcal{R}}_e^\infty \subseteq \bar{\mathcal{E}}_e$ for a given set of compromised sensors (i.e., a given $\Gamma$). Using this ellipsoidal approximation, obtaining $F$ can be formulated as a convex optimization problem by minimizing the impact of all bounded perturbations on the closed-loop system in \eqref{eq:DI_ClosedLoop_residue}. Note that $F$ must be computed for the same class of attacks as $\bar{\mathcal{E}}_e$, that is, the same $\Gamma$ must be considered. Optimizing $F$ by minimizing the reachable set induced by all bounded perturbations maximizes the resilience against stealthy attacks.

\subsection{Controller Realization Including Detector Scheme} \label{sec:CRIDS}
The closed-loop system in \eqref{eq:DI_ClosedLoop_residue}, restated below for the sake of readability, is subject to bounded perturbations $\omega \in \mathcal{E}_\omega, v \in \mathcal{E}_v, \bar e \in \bar{\mathcal{E}}_e$, and $r \in \mathcal{E}_r$:
\begin{multline} \nonumber
    \dot{\zeta} = \mathcal{A} \zeta + \mathcal{G} \omega + (\mathcal{H} - \mathcal{T} \Gamma^\dagger H) v  + \mathcal{T} \Gamma^\dagger r - \mathcal{T} \Gamma^\dagger \bar e_1,
\end{multline}
where $\mathcal{H}$ and $\mathcal{T}$ are affine in $F$. To enhance the robustness of the closed-loop system to the aforementioned perturbations, the goal is to find $F$ that minimizes the infinite-time reachable set $\mathcal{R}_\zeta^\infty \coloneqq \lim_{t \to \infty} {\mathcal{R}}_\zeta (t)$. The reachable set $\mathcal{R}_\zeta (t)$ defines all system states emanating from the initial condition $\zeta(0) = \zeta_0\in\mathbb{R}^{n_\rho}$, under all bounded perturbations, i.e., 
\begin{align} \label{eq:DI_RS_zeta}
    \! \! \! \mathcal{R}_\zeta (t) \! \coloneqq \! \left\{  \zeta(t)  \middle|
    \begin{aligned}
        & \exists\, \omega(s)\in \mathcal{E}_\omega, v(s) \in \mathcal{E}_v, \\
        & \bar e(t) \in \bar{\mathcal{E}}_e, r(t) \in \mathcal{E}_r, \text{ s.t. } \zeta(s)\\ 
        & \text{solution to } \eqref{eq:DI_ClosedLoop_residue}, s \in [0, t]
    \end{aligned}
    \right\}.
\end{align}
Rather than minimizing $\mathcal{R}^\infty_\zeta$ exactly, we seek to minimize an ellipsoidal outer approximation and introduce $\mathcal{E}_\zeta \coloneqq \{\zeta \in \mathbb{R}^{n_\zeta} \mid 1 - \zeta^\top P_\zeta \zeta \geq 0 \}$.
\begin{thm}[Optimal PEC Realization] \label{thm:findF} \
    Let the conditions of Lemma~\ref{lemma:ErrorSet} be satisfied and consider the corresponding estimation error ellipsoid $\bar{\mathcal{E}_e}$ induced by stealthy attacks with selection matrix $\Gamma$. Further, consider the perturbed closed-loop dynamics~\eqref{eq:DI_ClosedLoop_residue} and the reachable set \eqref{eq:DI_RS_zeta}. For a given $\alpha_\zeta \in \mathbb{R}_{\geq0}$ and $\Gamma$, if there exist matrices $F \in \mathbb{R}^{n_\rho \times n_y}$ and $Y_\zeta \in \mathbb{R}^{n_\zeta \times n_\zeta}$, and constants $\beta_{\zeta,\omega}, \beta_{\zeta,v}, \beta_{\zeta,e}, \beta_{\zeta,r} \in \mathbb{R}_{\geq0}$, being the solution to the convex program:
         \begin{subequations} \label{eq:thm_findF}
            \begin{equation}
                \left\{\!\begin{aligned}
                    &\min_{{Y}_{\zeta}, F, \beta_{\zeta,\omega}, \beta_{\zeta,v}, \beta_{\zeta,e}, \beta_{\zeta,r}}\trace [{Y}_{\zeta}],\\
            	&\text{\emph{s.t.}} \quad Y_{\zeta}\succ 0, \quad  \beta_{\zeta,\omega}, \beta_{\zeta,v}, \beta_{\zeta,e}, \beta_{\zeta,r} \in\mathbb{R}_{\geq0}, \\
                 & -\tilde{\mathcal{M}}_\zeta - \alpha_\zeta \tilde{\mathcal{N}}_\zeta - \beta_{\zeta,\omega} S_{\omega,\kappa} - \\
                 & \hspace{10mm} \beta_{\zeta,v} S_{v,\kappa} - \beta_{\zeta,e} S_{e,\kappa} - \beta_{\zeta,r} S_{r,\kappa} \succeq 0
                \end{aligned}\right.
            \end{equation}
    with matrices
            \begin{align}
                \tilde{\mathcal{N}}_{\zeta} &\coloneqq  \diag \begin{bmatrix}Y_\zeta, 0, 0, 0, 0, -1 \end{bmatrix}, \\
                \tilde{\mathcal{M}}_\zeta &= \begin{bmatrix}
                        Y_\zeta \mathcal A^\top + \mathcal A Y_\zeta & * & * & * & * & * \\
                        \mathcal{G}^\top  & 0 & * & * & * & * \\
                        (\mathcal{H}-\mathcal{T} \, \Gamma^\dagger H)^\top  & 0 & 0 & * & * & * \\
                        -\begin{bmatrix}\mathcal{T} \, \Gamma^\dagger & 0 \end{bmatrix}^\top & 0 & 0 & 0 & * & * \\
                        (\mathcal{T} \, \Gamma^\dagger)^\top & 0 & 0 & 0 & 0 & * \\
                        0 & 0 & 0 & 0 & 0 & 0
                        \end{bmatrix},\\
                S_{\omega,\kappa,} &\coloneqq  \diag \begin{bmatrix}0, -W_\omega, 0, 0,0, 1\end{bmatrix}, \\
                S_{v,\kappa} &\coloneqq  \diag \begin{bmatrix}0, 0, -W_v, 0,0, 1\end{bmatrix}, \\
                S_{e,\kappa} &\coloneqq  \diag \begin{bmatrix}0, 0, 0, -\bar{P}_e,0, 1\end{bmatrix}, \\
                S_{r,\kappa} &\coloneqq  \diag \begin{bmatrix}0, 0, 0, 0,-\Pi, 1\end{bmatrix}, 
            \end{align}    
        \end{subequations}
    then, for $P_\zeta = (Y^*_\zeta)^{-1}$ the ellipsoidal set 
    \begin{align} \label{eq:CR_zetaEllipse}
        \mathcal{E}_\zeta \coloneqq \{\zeta \in \mathbb{R}^{n_\zeta} \mid 1 - \zeta^\top P_\zeta \zeta \geq 0\}
    \end{align}
    is forward-invariant and contains the infinite-time reachable set ${\mathcal{R}}_\zeta^\infty \coloneqq \lim_{t \to \infty} {\mathcal{R}}_\zeta (t) \subseteq \mathcal{E}_\zeta$. Moreover, ${\mathcal{E}}_\zeta$ yields the smallest upper bound on the asymptotic volume among all outer ellipsoidal approximations of $\mathcal{R}_\zeta^\infty$. The corresponding optimal PEC realization matrix $F^*$ is the one that achieves this bound. In other words, $F^*$ defines the realization that minimizes the volume of $\mathcal{E}_\zeta$,  and hence characterizes the optimal PEC realization.
\end{thm}
\begin{pf} Define the Lyapunov function candidate 
\begin{align} \label{eq:DI_Vzeta_att} 
    V(\zeta) &= \zeta^\top P_\zeta \zeta \geq 0, \quad P_\zeta \succ 0,
\end{align}
with $ P_\zeta \in \mathbb{R}^{n_\zeta \times n_\zeta}$. For the ellipsoidal set $ \mathcal{E}_\zeta$ in \eqref{eq:CR_zetaEllipse} to be a forward invariant set, we require that~\citep{escudero_analysis_2022,boyd_linear_1994}
\begin{subequations} \label{eq:DI_dVzeta_att1} \begin{align} 
     \dot{V}(\zeta, \omega, v, \bar e, r) = \dot{\zeta}^\top P_\zeta \zeta + \zeta^\top P_\zeta \dot{\zeta} \leq 0,
\end{align}
for all $\zeta$ such that ${V}(\zeta) \geq 1$, and all admissible inputs $\omega \in \mathcal{E}_w, v\in \mathcal{E}_v, \bar e \in \bar{\mathcal{E}}_e$, and $r\in\mathcal{E}_r$.
Substituting the dynamics from~\eqref{eq:DI_ClosedLoop_residue} yields
\begin{multline} \label{eq:DI_dVzeta_att2} 
        \! \! \! \dot{V}(\zeta, \omega, v, \bar e, r) = \left(\mathcal{A} \zeta + \mathcal{G} \omega + (\mathcal{H}-\mathcal{T} \, \Gamma^\dagger H) v +  \mathcal{T} \, \Gamma^\dagger r \right. \\
        \left. - \begin{bmatrix} \mathcal{T} \, \Gamma^\dagger & 0 \end{bmatrix} \bar e\right)^\top P_\zeta \zeta  
        + \zeta^\top P_\zeta (\mathcal{A} \zeta + \mathcal{G} \omega \\ + (\mathcal{H}-\mathcal{T} \, \Gamma^\dagger H) v + \mathcal{T} \, \Gamma^\dagger r - \begin{bmatrix} \mathcal{T} \, \Gamma^\dagger & 0 \end{bmatrix} \bar e) \leq 0.
\end{multline}
\end{subequations}
Define stacked vector $\kappa = \begin{bmatrix} \zeta & \omega & v & \bar e & r & 1 \end{bmatrix}^\top$,  which allows the reformulation of~\eqref{eq:DI_Vzeta_att}, \eqref{eq:DI_dVzeta_att1}, $\mathcal{E}_\omega$, $\mathcal{E}_v$, $\bar{\mathcal{E}}_e$ and $\mathcal{E}_r$ as
\begin{subequations} \label{eq:DI_InputSet_kappa} \begin{align}
    V_\kappa( \kappa) &= \kappa^\top \mathcal{N}_\zeta \kappa, \quad
    \dot{V}_\kappa(\kappa) = \kappa^\top \mathcal{M}_\zeta \kappa, \\
    \! \! \! \! \mathcal{E}_\omega \! & \coloneqq \! \{\omega \in \mathbb{R}^{n_x} \! \mid \! {\kappa}^\top \mathcal{S}_{\omega,\kappa} {\kappa} \geq 0,  &&\!\!\!\!\!\!\!\! \forall \zeta,\bar e,v,r\}, \\
    \! \! \! \!  \mathcal{E}_v \!&\coloneqq \! \{v \in \mathbb{R}^{n_v} \! \mid \! {\kappa}^\top \mathcal{S}_{v,{\kappa}} {\kappa} \geq 0, &&\!\!\!\!\!\!\!\! \forall \zeta, \bar e,\omega,r \}, \\
    \! \! \! \!  \bar{\mathcal{E}}_e \!&\coloneqq \! \{\bar e \in \mathbb{R}^{n_x} \! \mid \! {\kappa}^\top \mathcal{S}_{e,\kappa} {\kappa} \geq 0,  &&\!\!\!\!\!\!\!\! \forall \zeta, \omega,v,r \},\\
    \! \! \! \!  \mathcal{E}_r \!&\coloneqq \! \{r \in \mathbb{R}^{n_y} \! \mid \! {\kappa}^\top \mathcal{S}_{r,\kappa} {\kappa} \geq 0,  &&\!\!\!\!\!\!\!\! \forall \zeta, \bar e,\omega,v \}
\end{align} 
with matrices
    \begin{align}
         {\mathcal{N}}_{\zeta} &= \diag \begin{bmatrix}P_\zeta, 0, 0, 0, 0, -1 \end{bmatrix}, \\
        {\mathcal{M}}_\zeta &= 
        \begin{bmatrix}
                 \mathcal A^\top P_\zeta + P_\zeta \mathcal A  & * & * & * & * & * \\
                \mathcal{G}^\top P_\zeta & 0 & * & * & * & * \\
                (\mathcal{H}-\mathcal{T} \, \Gamma^\dagger H)^\top P_\zeta & 0 & 0 & * & * & * \\
                -\begin{bmatrix}\mathcal{T} \, \Gamma^\dagger  & 0
                \end{bmatrix}^\top P_\zeta & 0 & 0 & 0 & * & * \\
                (\mathcal{T} \, \Gamma^\dagger)^\top P_\zeta& 0 & 0 & 0 & 0 & * \\
                0 & 0 & 0 & 0 & 0 & 0
        \end{bmatrix}, 
    \end{align}
\end{subequations}
and $S_{\omega,\kappa}, S_{v,\kappa}, S_{e,\kappa}, S_{r,\kappa}$ as in~\eqref{eq:thm_findF}. By the S-procedure \citep{boyd_linear_1994}, the inequalities in~\eqref{eq:DI_Vzeta_att}--\eqref{eq:DI_InputSet_kappa} are implied if multipliers ${\alpha}_\zeta, {\beta}_{\zeta,\omega}, {\beta}_{\zeta,v}, {\beta}_{\zeta,e}, {\beta}_{\zeta,r}\in \mathbb{R}_{\geq0}$ exist such that 
    \begin{multline} \label{eq:Proposition_F_nonlinear}
        -{\mathcal M}_{\zeta} - {\alpha}_{\zeta} {\mathcal N}_{\zeta} - {\beta}_{\zeta,\omega} \mathcal{S}_{\omega, \kappa} - {\beta}_{\zeta,v} \mathcal{S}_{v, \kappa} \\ 
                 - {\beta}_{\zeta,e} \mathcal{S}_{e, \kappa} - {\beta}_{\zeta,r} \mathcal{S}_{r,\kappa}  \succeq 0. 
    \end{multline}
The resulting inequality in~\eqref{eq:Proposition_F_nonlinear} is nonlinear in the unknown $F$ and $P_\zeta$ due to the cross product of matrices $\mathcal{T}$, $\mathcal{H}$ (both depending on $F$) with $P_\zeta$ in $\mathcal{M}_\zeta$. Therefore, a congruence transformation of the form $Q W Q^\top \succeq 0$ is applied where $W$ is the left-hand side in \eqref{eq:Proposition_F_nonlinear}, and $Q = Q^\top = \diag \begin{bmatrix}
Y_\zeta & I& I& I&1\end{bmatrix}$, with $Y_\zeta =P^{-1}_\zeta$, which preserves the definiteness of the matrix inequality~\citep{boyd_linear_1994}. As a result, we obtain the inequality 
\begin{multline} \label{eq:Proposition_F_linear}
    -\tilde{\mathcal M}_{\zeta} - {\alpha}_{\zeta} \tilde{\mathcal N}_{\zeta} - {\beta}_{\zeta,\omega} \mathcal{S}_{\omega, \kappa} - {\beta}_{\zeta,v} \mathcal{S}_{v, \kappa} \\  - {\beta}_{\zeta,e} \mathcal{S}_{e, \kappa} - {\beta}_{\zeta,r} \mathcal{S}_{r,\kappa}  \succeq 0 
\end{multline}
with $\tilde{\mathcal{M}}_\zeta$ and $\tilde{\mathcal{N}}_\zeta$ as in \eqref{eq:thm_findF}. Note that $S_{\omega,\kappa},S_{v,\kappa},S_{e,\kappa},S_{r,\kappa}$ remain unaffected due to the choice of $Q$. 

To ensure that the ellipsoidal bound is as tight as possible, we minimize $-\log\det(P_\zeta)$, which is proportional to the volume of the ellipsoid~\citep{boyd_linear_1994}. However, after applying the congruence transformation, the objective function $\min -\log\det[Y_\zeta]$ becomes concave. Therefore, we instead minimize a convex upper bound on the ellipsoidal volume by minimizing $\trace[Y_\zeta]$~\citep{murguia_security_2020}. Feasibility of the convex program implies that the ellipsoid ${\mathcal{E}}_\zeta$ is forward-invariant and, among all ellipsoids containing the asymptotic reachable set ${\mathcal{R}}_\zeta^\infty \coloneqq \lim_{t \to \infty} {\mathcal{R}}_\zeta (t)$, provides the smallest known volume upper bound. Finally, the corresponding matrix $F^*$ defines the optimal PEC realization in the sense that it minimizes this upper bound.
 \null\hfill $\blacksquare$
\end{pf}
\begin{rem}[Optimality of the Base Controller] \
Under Assumption~\ref{assum:ExistenceDOFC}, the base controller $\mathcal{F}$ ensures that the nominal closed-loop system is stable, i.e., $\mathcal{A}$ in~\eqref{eq:thm_findF} is Hurwitz. Consequently, the Lyapunov conditions underlying the convex program in Theorem~\ref{thm:findF} admit a feasible solution. In certain configurations, the optimal solution may yield $F^* = 0$. In this case, the base controller already represents the optimal PEC realization, and introducing PEC realizations cannot further reduce the volume upper bound of $\mathcal{E}_\zeta$ for the considered attack selection matrix $\Gamma$.
\end{rem}

\section{Case Study Results} \label{sec:RESULTS}
This section presents the application of the developed methods to the quadruple-tank process using the decentralized PI controller from~\citep{johansson_quadruple_tank_2000} as a base controller.

\begin{figure}[bt]
    \centering
    \resizebox{0.9\columnwidth}{!}{\input{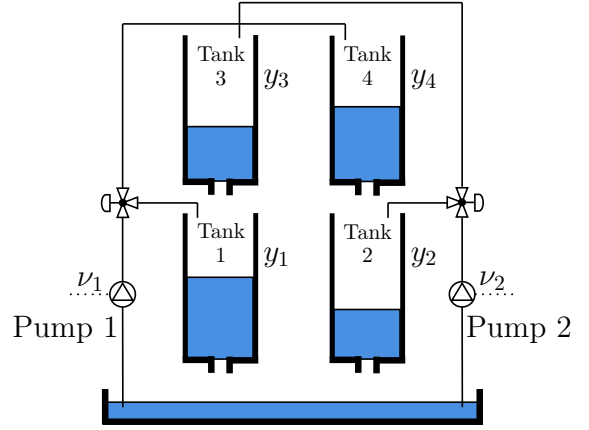}}
    \caption{Schematic overview of the quadruple-tank process. The water levels in tanks 1 and 2 are controlled using two pumps with voltage inputs $\nu_j, \, j \in \{1,2\}$, using the measured tank levels $y_i, \, i \in \{1,2,3,4\}$.}
    \label{fig:RS_quadrupleTank}
\end{figure}

A schematic of the system is shown in Fig.~\ref{fig:RS_quadrupleTank}. The process aims to control the water levels $h_1$ and $h_2$ in the lower tanks using two pumps. The input voltages to the pump $\nu_j$, $j\in\{1.2\}$, generate flows $k_j \nu_j$ with pump constant $k_j$. The measured outputs $y_i$ denote the tank levels $y_i = k_c h_i, i\in\{1,2,3,4\}$, with sensor parameter $k_c$. Unlike~\citep{johansson_quadruple_tank_2000}, we assume to have water level measurements from all four tanks, which is necessary to ensure the existence of a realization matrix $F$ satisfying~Lemma~\ref{lem:ExistsenceF}. 

The system is linearized around the operating point $(h_i^0, \nu_j^0)$, with deviations defined as $x_i \coloneqq h_i - h_i^0$ and $u_j \coloneqq \nu_j - \nu_j^0$. Although full state information is available, we define the transformed state $\bar{x} = Cx$ with state-space model
\begin{align}\label{eq:RS_stateSpace}\begin{aligned}
    \dot{\bar x} &= \bar A \bar x + \bar B u + \bar G \omega,  \quad
    \tilde y = \bar C \bar{x} + H v + \Gamma \delta_y, \\
    \bar{A} &=\!\! 
    \left[\begin{smallmatrix}
        \tfrac{-1}{T_1} & 0 & \tfrac{A_3}{A_1 T_3} & 0 \\
        0 & \tfrac{-1}{T_2} & 0 & \tfrac{A_4}{A_2 T_4} \\
        0 & 0 & \tfrac{-1}{T_3} & 0\\
        0 & 0 & 0 & \tfrac{-1}{T_4}
    \end{smallmatrix}\right] \!\!, \bar B \!=\!\!
    \left[ \begin{smallmatrix}
        \tfrac{\gamma_1 k_1 k_c}{A_1} & 0 \\
        0 & \tfrac{\gamma_2 k_2 k_c}{A_2} \\
        0 & \tfrac{\gamma_2' k_2 k_c}{A_3} \\
        \tfrac{\gamma_1' k_1 k_c}{A_4} & 0 
    \end{smallmatrix}\right]\!\!, \\
    \bar{G} &=\bar{B}, \quad \bar{C} = H = I_4.
\end{aligned}\end{align}

The time constants are $T_i = \tfrac{A_i}{a_i} \sqrt{\tfrac{2h_i^0}{g}}$, where $A_i$ and $a_i$ denote the cross-sectional area of tank $i\in\{1,2,3,4\}$ and its outlet hole, respectively. The valve settings $\gamma_j\in(0,1), \, j\in\{1,2\}$ determine the flow split, with $\gamma_j'=(1-\gamma_j)$, and $g$ is the gravitational constant. All the system parameters are given in Table~\ref{tab:SysParam}.  
\begin{table}[bt]
    \centering  
    \caption{System parameters for the quadruple-tank process depicted in Figure~\ref{fig:RS_quadrupleTank}, with its state space formulation in~\eqref{eq:RS_stateSpace}~\citep{johansson_quadruple_tank_2000}.}
    \resizebox{\columnwidth}{!}{%
    \begin{tabular}{l | l | l } \hline\hline
    $(A_1,A_2,A_3,A_4)$ & [cm$^2$] & $(28,32,28,32)$ \\ 
    $(a_1,a_2,a_3,a_4)$ & [cm$^2$] &$(7.1,5.7,7.1,5.7)e^{-2}$ \\ 
    $(h^0_1,h^0_2,h^0_3,h^0_4)$ & [cm] &$(12.4,12.7,1.8,1.4)$ \\ 
    $(\nu^0_1,\nu^0_2)$ & [V] &$(3.00,3.00)$ \\ 
    $(k_1,k_2)$ & [cm$^3$/Vs] &$(3.33,3.35)$ \\ 
    $(\gamma_1,\gamma_2)$ & [-] &$(0.70,0.60)$ \\ 
    $k_c$ & [V/s$^2$] &$0.50$ \\ 
    $g$ & [cm/s$^2$] &$981$ \\ 
    $(\bar v, \bar \omega)$ & [V] & $(0.05,0.003)$
    \end{tabular}} \label{tab:SysParam}
\end{table}
The desired closed-loop behavior is achieved by adopting the decentralized PI from~\citep{johansson_quadruple_tank_2000} as the base controller. By defining the integrator states $\rho_j~=~\int_0^t y_{r,j}(\tau)- \tilde y_j(\tau) d\tau, \, j \in \{1,2\}$, we obtain the following base controller:
\begin{align} \label{eq:RS_baseDOFC} \mathcal{F} = \left \{\begin{aligned}
        \dot{\rho}_j &= y_{r,j}-\tilde y_j,\\
        u_j &= 
        \tfrac{K_j}{T_{Ij}} 
        \rho_j + 
        K_j  (y_{r,j}-\tilde y_j)
\end{aligned}\right.\end{align}
with controller settings $(K_1,T_{I1}) = (3.0,30)$ and $(K_2,T_{I2}) = (2.7,40)$.

To approximate the residual set $\mathcal{E}_r$ and estimation error set $\bar{\mathcal{E}}_e$ via Lemma~\ref{lemma:StealthySet} and Lemma~\ref{lemma:ErrorSet}, respectively, a detector scheme $\mathcal{D}$ is required. As designing optimal detector schemes has been studied in prior work, the objective here is not to co-design the detector but to robustify a given base controller under the assumption that detection fails, i.e., during a stealthy FDI attack. Therefore, given that $(\bar A,\bar C)$ is observable, we select $\bar L$ arbitrarily such that the eigenvalues of $(\bar A - \bar L \bar C)$ are placed at $[-2.0,-2.0,-2.1-2.1]$, which yields a residual $\mathcal{E}_r$ set with $-\log\det(\Pi)=-19.26$. 

For each sensor attack scenario $s$ (and corresponding $\Gamma$), the error set $\bar{\mathcal{E}}_e$ is computed, and afterward the corresponding optimal realization $F_s^*$ is obtained via Theorem~\ref{thm:findF}. The considered use-cases include three attack scenarios: (i) a single sensor measurement is compromised, (ii) simultaneous attack on two sensors, either both measuring tanks actuated by the same pump or both located on the same side of the system (which are dynamically coupled), (iii) all sensors are compromised. The results are summarized in Table~\ref{tab:Results}, which were computed using MATLAB 2022b and MOSEK 11.3, and can be obtained via~\citep{huisman2025_github}. 
\begin{table}[bt]
    \centering
    \caption{Cost values associated with the approximation $\bar{\mathcal{E}}e$ of the reachable set $\bar{\mathcal{R}}e^\infty$ for various sensor attacks, computed using Lemma~\ref{lemma:ErrorSet}. Additionally, the table shows the cost values from Theorem~\ref{thm:findF}, comparing the base controller ($F = \mathbf{0}$, with cost $\trace(Y_\zeta)$) and the optimized realization ($F^*$, with cost $\trace(Y_\zeta^*)$).}
    \label{tab:Results}
     \resizebox{\columnwidth}{!}{%
    \begin{tabular}{c|c|c|c} \hline\hline
         Att. Sensors & $-\log\det\left(\bar P_e\right)$ & $\trace(Y_\zeta)$ & $\trace(Y^*_\zeta)$ \\ \hline
         $\{1\}$ & $-11.52$ & $85360$ & $75785$ \\
         $\{4\}$ & $-13.66$ & $1.28$ & $1.27$ \\
         $\{1,4\}$ & $8.66$ & $85365$ & $75793$ \\
         $\{2,4\}$ & $-4.43$ & $561965$ & $477978$ \\
         $\{1,2,3,4\}$ & $23.66$ & $1326004$& $1115141$ \\
    \end{tabular}}
\end{table}

Notably, attacks on upper-tank sensors (e.g., $y_4$) result in minimal degradation, as indicated by the cost function values. In particular, since the base controller does not utilize $y_4$, this sensor is effectively redundant. The optimal realization under attack on $y_4$ confirms this, as it reconfigures the controller to rely on the remaining healthy sensors $y_1$, $y_2$, and $y_3$ to improve robustness:
\begin{align}
    F^*_{\{4\}} &= \begin{bmatrix}
       -1.04 & 0.00 & -1.60 & 0.00 \\
       -1.84 & 0.00 & -3.30 & 0.00
    \end{bmatrix}.
\end{align}
A similar trend is observed when both $y_1$ and $y_4$ are compromised. The corresponding optimal realization,
\begin{align}
    F^*_{\{1,4\}} &= \begin{bmatrix}
       -3.84 & 0.00 & -5.93 & 0.00 \\
       -6.32 & 0.00 & -11.36 & 0.00
    \end{bmatrix},
\end{align}
demonstrates that the algorithm effectively decouples the influence of $y_4$, while mitigating the impact of the compromised $y_1$ by leveraging $y_3$. 

The resulting matrix $F^*$, with the corresponding cost in Table~\ref{tab:Results}, highlight the redundancy and relative criticality of different sensors. The values for $\trace(Y_\zeta)$ indicate that an attack on $y_1$ and $y_2$ causes significant damage to the system, compared to an attack on $y_4$ (the same conclusion can be drawn for $y_3$, as it mirrors an attack on $y_4$). Moreover, compromising the sensors of tanks that are controlled by the same pump (i.e., tanks 1 and 4) has a smaller impact than compromising the sensors of the dynamically coupled tanks (i.e., tanks 2 and 4). In conclusion, the proposed framework not only enhances robustness under sensor attacks but also provides insight into sensor criticality, aiding in security-aware design and sensor protection strategies.

To demonstrate the performance of an optimized controller realization, a simulation study is conducted for an FDI attack on $y_1$, as this sensor is essential for the base controller. Using Theorem~\ref{thm:findF} we obtain 
\begin{align}
    F^*_{\{1\}} &= \begin{bmatrix}
        -3.84 & -1.10 & -5.93 & 0.12 \\
        -6.32 & -1.60 & -11.37 & -2.73
    \end{bmatrix},
\end{align}
which achieves the reduction in $\trace(Y_\zeta)$ reported in Table~\ref{tab:Results}. The model \eqref{eq:RS_stateSpace} is implemented in MATLAB/Simulink, including the detector introduced above. The simulation code can be found in~\citep{huisman2025_github}.

To ensure stealthiness, the attack $\delta_y$ is constructed according to~\eqref{eq:DI_attackSignal} and includes a small sinusoidal perturbation:
\begin{align}\label{eq:RS_FDI}
    \delta_y = \begin{bmatrix}
        1 & 0 & 0 & 0
    \end{bmatrix} \left (r_1 - e_1 - v_1 \right) + 0.1\sin{(0.25t)},
\end{align}
where $r_1$, $e_1$, and $v_1$ denote, respectively, the residual, the estimation error, and sensor noise associated with output $y_1$. The sinusoidal perturbation is chosen arbitrarily, and $\delta_y$ satisfies $r^ \top \Pi r \leq 1$ and therefore remains undetected. 

\begin{figure}[bt]\centering
	\includegraphics[width=0.85\linewidth]{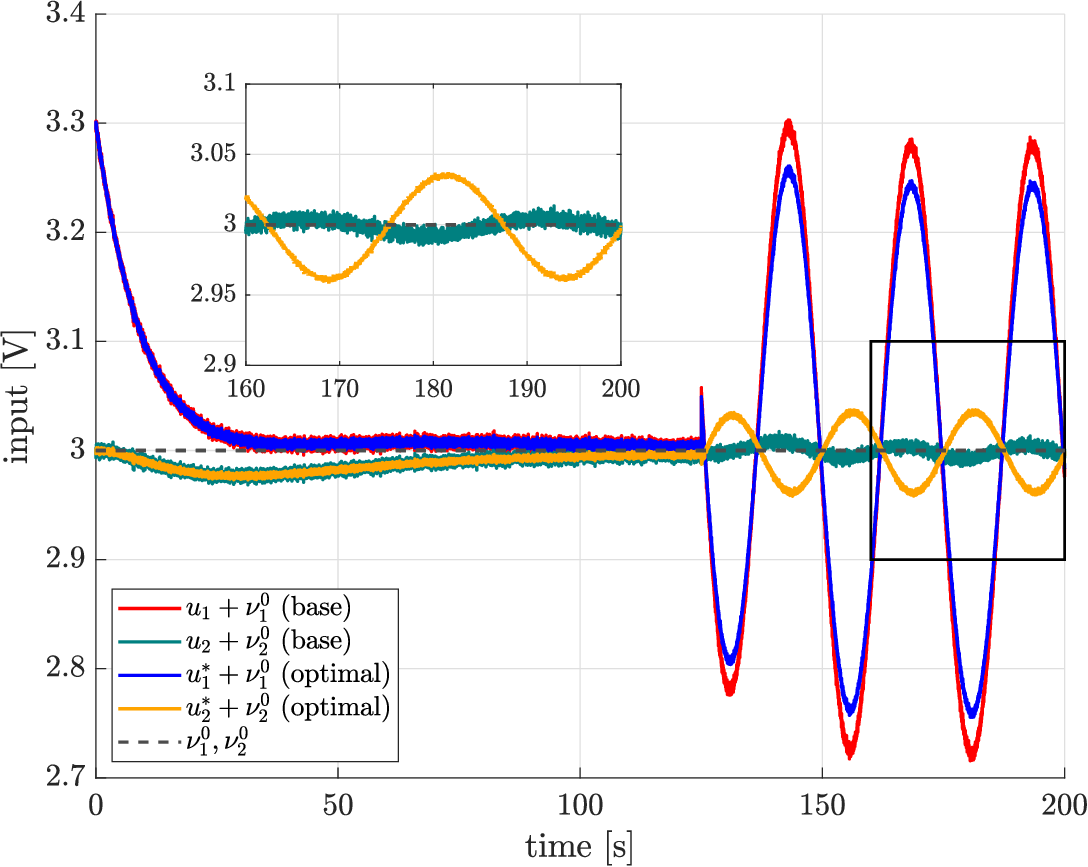}
        \caption{Behavior of the base controller $u_j$ and the optimized controller realization $u_j^*$, for $j \in \{1,2\}$, with false data injection \eqref{eq:RS_FDI} at $t = 125\,\mathrm{s}$. The optimized controller enhances robustness against the attack for $u_1^*$, while reducing the robustness of $u_2^*$ compared to $u_2$. Notably, $u_j^*$ achieves improved noise attenuation relative to the base controller.}
	\centering
	\label{fig:RS_Controller}
\end{figure}

\begin{figure}[bt]\centering
	\includegraphics[width=\linewidth]{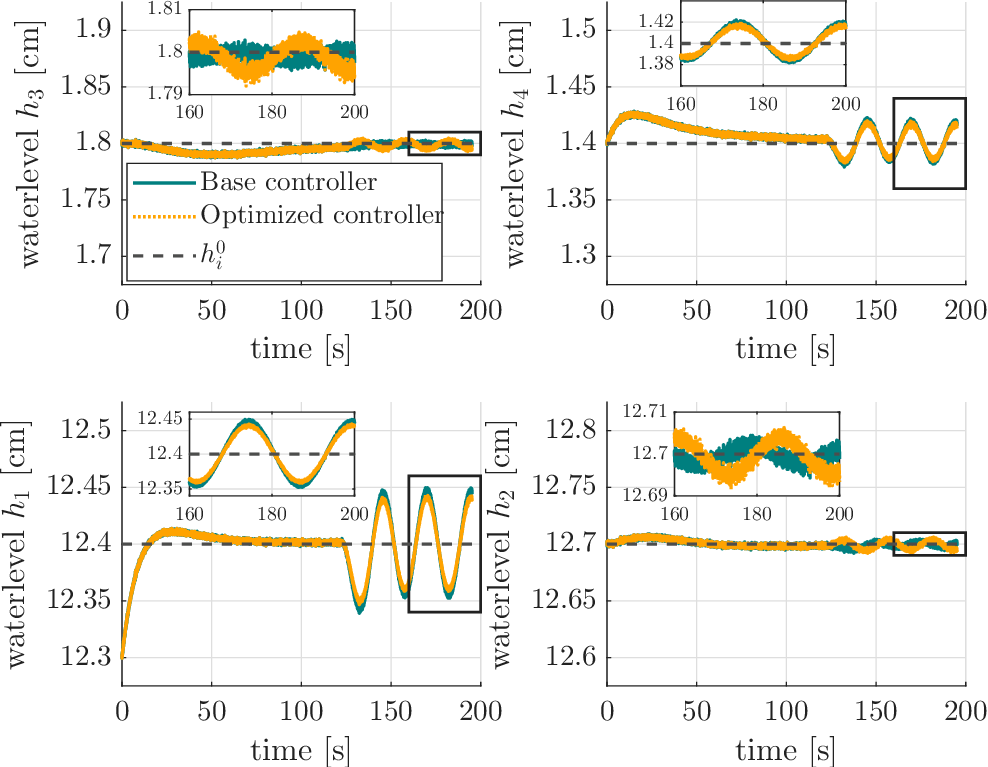}
        \caption{Closed-loop responses of the quadruple-tank process using the base controller $u_j$ and the optimized controller realization $u_j^*$, for $j \in \{1,2\}$, with false data injection \eqref{eq:RS_FDI} at $t = 125\,\mathrm{s}$. Each subplot depicts the water level $h_i$ of tank $i \in \{1,2,3,4\}$, showing improved robustness in tanks $h_1$ and $h_4$, while reducing it for tanks $h_2$ and $h_3$.}
	\centering
	\label{fig:RS_Tanks}
\end{figure}

Figures~\ref{fig:RS_Controller} and~\ref{fig:RS_Tanks} depict the resulting tank-level and controller responses under an FDI attack for $t \geq 125\,\mathrm{s}$. For $\delta_y = 0$, the optimized realization $u_j^*$, $j\in\{1,2\}$ exhibits the same nominal performance as the base controller $u_j$, with slightly improved noise attenuation, as evidenced by smoother control profiles. For $\delta_y \neq 0$, the optimized realization $u_j^*$, enhances the robustness of tanks $h_1$ and $h_4$, while a degradation in performance appears in $h_2$ and $h_3$, reflecting a robustness–performance trade-off. The control inputs confirm this trend: while $u_1^*$ effectively suppresses the FDI attack compared to $u_1$, $u_2^*$ exhibits slightly more amplification than $u_2$. Overall, the simulation results align with the findings in Table~\ref{tab:Results}, demonstrating that the proposed method enhances the system’s resilience when the FDI attack on sensor $y_1$ is introduced at $t = 125\,\mathrm{s}$. The optimized realization effectively reduces the impact on the most affected tanks $h_1$ and $h_4$, while causing only a small degradation in performance observed for $h_2$ and $h_3$. 

\begin{comment}
Figures~\ref{fig:RS_Controller} and~\ref{fig:RS_Tanks} depict the resulting tank-level and controller responses under an FDI attack for $t \geq 125\,\mathrm{s}$. The optimized realization $u_j^*$, $j\in\{1,2\}$, improves robustness of tanks $h_1$ and $h_4$, while a degradation in performance appears in $h_2$ and $h_3$, reflecting a robustness–performance trade-off. The control inputs confirm this trend: while $u_1^*$ effectively suppresses the FDI attack compared to $u_1$, $u_2^*$ exhibits slightly more amplification than $u_2$. Furthermore, $u_j^*$ achieves improved noise attenuation, as evidenced by the smoother control profiles. Overall, the simulation results align with the findings in Table~\ref{tab:Results}, demonstrating that the proposed method enhances the system’s resilience when the FDI attack on sensor $y_1$ is introduced at $t = 125\,\mathrm{s}$. The optimized realization effectively reduces the impact on the most affected tanks $h_1$ and $h_4$, while causing only a small degradation in performance observed for $h_2$ and $h_3$. 

\end{comment}

%% Optimization Formulation
\section{Conclusion and Future Research} \label{sec:Conclusion}
This paper proposed a controller-oriented framework to enhance the resiliency of cyber-physical systems (CPSs) against stealthy false data injection (FDI) attacks without compromising nominal closed-loop performance. By reformulating a dynamic output-feedback controller, a class of \emph{equivalent controller realizations} was derived that preserves nominal input–output behavior while exhibiting different robustness properties under disturbances and attacks. Reachable set analysis was employed to quantify the effect of disturbances and stealthy attacks, and an LMI problem was formulated to compute the \emph{optimal controller realization} minimizing the reachable set. The effectiveness of the approach was demonstrated on the quadruple-tank process.

Future research will focus on extending the concept of PEC realizations towards attack detection and mitigation, and on steering the system states towards a predefined safe set, i.e., ensuring that trajectories remain within this set, rather than solely minimizing the overall reachable set. In particular, PEC realizations may provide a promising framework for detecting and mitigating actuator attacks by exploiting structural redundancies in controller implementations.

%% Referencelist
\bibliographystyle{apalike}        % Include this if you use bibtex 
\bibliography{referenceList}    % and a bib file to produce the 
                                 % bibliography (preferred). The
                                 % correct style is generated by
                                 % Elsevier at the time of printing.

%\printbibliography

\appendix
\section{Stealthy Attack Set}    % Each appendix must have a short title.
\label{app:obtianPi}
This appendix provides the tools to obtain the residual set from Section~\ref{sec:GOFD}. The objective is to quantify the class of stealthy attacks, i.e., perturbations $\delta_y$, such that the residual $r$ remains within the ellipsoidal set $\mathcal{E}_r\coloneqq \{ r \in \mathbb{R}^{n_y} \mid 1- r^\top \Pi r \geq 0 \}$, thereby avoiding detection. Consider the error dynamics in \eqref{eq:SA_ErrorDyn1} in the absence of an attack (i.e., $\delta_y = 0$):
\begin{equation}\label{eq:DI_errorDyn_attFree}
\left\{
\begin{aligned}
    \dot{e} &= A_e e + G \omega - L H v, \\
    r &= C e + H v,
\end{aligned}
\right.
\end{equation}
with $A_e = (A - LC)$ and disturbances $\omega \in \mathcal{E}_\omega$, $v \in \mathcal{E}_v$, as defined in~\eqref{eq:SA_inputSet_original}. The goal is to (i) quantify the reachable set ${\mathcal{R}}_e$ (Definition~\ref{defn:reachableSet}), and (ii) determine $\Pi$ such that $r \in \mathcal{E}_r$ for all $e \in {\mathcal{R}}_e$ and $v \in \mathcal{E}_v$.

The reachable set ${\mathcal{R}}_e$ contains all estimation error trajectories $e$ that originate from the initial condition $e(0) = e_0\in \mathbb{R}^{n_x}$, subject to $\omega \in \mathcal{E}_\omega$ and $v \in \mathcal{E}_v$, i.e., 
\begin{align} \label{eq:DI_RS_eStealthy}
    \!\!\!\!\mathcal{R}_e(t)\! \coloneqq \! \left\{ e(t) \middle|
    \begin{aligned}
        &\exists\,\omega(s) \in \mathcal{E}_\omega, v(s) \in \mathcal{E}_v, \text{ s.t. } e(s)\\
        &  \text{solution to } \eqref{eq:DI_errorDyn_attFree},\, s \in [0,t] 
    \end{aligned}
    \right\}.
\end{align}
Since exact computation of $\mathcal{R}_e$ is intractable, we seek an outer ellipsoidal approximation $\mathcal{E}_e \coloneqq \{ e \in \mathbb{R}^{n_x} \mid 1 - e^\top P_e e \geq 0 \}$, with positive definite matrix $ P_e \! \in \! \mathbb{R}^{n_x \times n_x}$, such that $\mathcal{R}_e^\infty \subseteq \mathcal{E}_e$, where $\mathcal{R}_e^\infty = \lim_{t \to \infty} \mathcal{R}_e(t)$.  
\begin{lem}[Residual Set] \label{lemma:StealthySet} \
    Consider the dynamics~\eqref{eq:DI_errorDyn_attFree} and reachable set~\eqref{eq:DI_RS_eStealthy}. Given constants $\alpha_e, \alpha_r \in \mathbb{R}_{\geq0}$, if there exist $P_e \in \mathbb{R}^{n_x \times n_x}$, $\Pi \in \mathbb{R}^{n_y \times n_y}$, and $\beta_{e,\omega}, \beta_{e,v}, \beta_{r,v} \in \mathbb{R}_{\geq0}$, being the solution to the convex program:
         \begin{subequations} \label{eq:Proposition_stealthySet}
            \begin{equation} \label{eq:Proposition_stealthySet1}
                \left\{\!\begin{aligned} 
                    &\min_{P_e, \Pi, \beta_{e,\omega}, \beta_{e,v}, \beta_{r,v}} - \log\det(P_e) - \log\det(\Pi),\\
            	&\text{\emph{s.t.}} \quad  P_{e}, \Pi\succ 0, \quad  {\beta}_{e,\omega}, {\beta}_{e,v}, {\beta}_{r,v}\in\mathbb{R}_{\geq0}, \\
                    &-\mathcal M_e - \alpha_e \mathcal N_e - \beta_{e,\omega} \mathcal S_{\omega,\kappa} - \beta_{e,v} \mathcal S_{v,\kappa}  \succeq 0,  \\
                    & -\mathcal M_{r} + \alpha_r \mathcal N_e - \beta_{r,v} \mathcal S_{v,\kappa}  \succeq 0, 
                \end{aligned}\right.
            \end{equation}
    with matrices
            \begin{align}
               \mathcal{N}_e  &\coloneqq
                \diag \begin{bmatrix}
                    P_e, 0, 0, -1  \\
                \end{bmatrix}, \label{eq:DI_Ne_attfree}\\
                \mathcal{M}_e  &\coloneqq 
                \begin{bmatrix}
                    A_e^\top P_e + P_e A_e & P_e G & - P_eLH & 0 \\
                    G^\top P_e & 0 & 0 & 0  \\
                    -(LH)^\top P_e & 0 & 0 & 0  \\
                    0 & 0 & 0 & 0  \\
                \end{bmatrix}, \\
                \mathcal{M}_{r} &\coloneqq \begin{bmatrix}
                    C^\top \Pi C & 0 & C^\top \Pi H & 0 \\
                    0 & 0 & 0 & 0 \\
                    H^\top \Pi C & 0 & H^\top \Pi H & 0 \\
                    0 & 0 & 0 & -1
                \end{bmatrix}, \label{eq:Proposition_stealthySet4} \\
                \mathcal{S}_{\omega,\kappa} &\coloneqq \diag \begin{bmatrix} 0, -W_\omega, 0, 1 \end{bmatrix} ,  \\
                \mathcal{S}_{v,\kappa} &\coloneqq \diag \begin{bmatrix} 0, 0, -W_v, 1 \end{bmatrix};
            \end{align}    
        \end{subequations}
    then the residual satisfies $r\in\mathcal{E}_r$ for all $e \in \mathcal{E}_e, \omega \in \mathcal{E}_\omega, v\in\mathcal{E}_v$. Moreover, the ellipsoidal set 
    \begin{align} \label{eq:DI_errorSet_Stealthy}
        {\mathcal{E}}_e := \left\{ {e} \in \mathbb{R}^{n_x} \mid 1- {e}^\top {P}_{e} {e} \geq 0 \right\}
    \end{align}
    is forward-invariant and contains the infinite-time reachable set ${\mathcal{R}}_e^\infty \coloneqq \lim_{t \to \infty} {\mathcal{R}}_e (t) \subseteq \mathcal{E}_e$. Moreover, the pair $(P_e,\Pi)$ returned by \eqref{eq:Proposition_stealthySet} minimizes the asymptotic volumes of the error ellipsoid $\mathcal{E}_e$ and the residual ellipsoid $\mathcal{E}_r\coloneqq \{ r \in \mathbb{R}^{n_y} \mid 1- r^\top \Pi r \geq 0 \}$ among all feasible solutions.
\end{lem}
\begin{pf} Define the Lyapunov function candidate 
\begin{align} \label{eq:DI_Ve_stealthy} 
     V(e) =  e^\top {P}_e  e \geq 0, \quad  P_e \succ 0,
\end{align}
with $ P_e \in \mathbb{R}^{n_x \times n_x}$. For $\mathcal{E}_e$ in \eqref{eq:DI_errorSet_Stealthy} to be forward invariant, we require that~\citep{escudero_analysis_2022,boyd_linear_1994}
\begin{align} \label{eq:DI_dVe_stealthy1}
    \dot{{V}}( e, \omega, v) = \dot{e}^\top  {P}_e e + e^\top {P}_e \dot{e} \leq 0,
\end{align}
when ${V}(e) \geq 1$, and $\omega \in \mathcal{E}_w, v\in \mathcal{E}_v$, with $\dot{e}$ as in~\eqref{eq:DI_errorDyn_attFree}.

Define the stacked vector ${\kappa} = \begin{bmatrix}  e & \omega & v & 1 \end{bmatrix}^\top$, which allows the reformulation of~\eqref{eq:DI_Ve_stealthy}, \eqref{eq:DI_dVe_stealthy1}, $\mathcal{E}_\omega$, and $\mathcal{E}_v$ as 
\begin{subequations} \label{eq:DI_InputSet_stealthy} \begin{align}
     V_\kappa( \kappa) &=  \kappa^\top {\mathcal{N}}_e  \kappa, \quad
    \dot{V}_\kappa( \kappa) =  \kappa^\top {\mathcal{M}}_e  \kappa, \\
    \! \! \! \mathcal{E}_\omega & \coloneqq \! \{\omega \in \mathbb{R}^{n_x} \! \mid \! {\kappa}^\top \mathcal{S}_{\omega,\kappa} {\kappa} \geq 0,  &&\!\!\!\! \forall  e,v\}, \\
    \! \! \!  \mathcal{E}_v &\coloneqq \! \{v \in \mathbb{R}^{n_v} \! \mid \! {\kappa}^\top \mathcal{S}_{v,{\kappa}} {\kappa} \geq 0, &&\!\!\!\!\forall  e,\omega \} 
\end{align} \end{subequations}
with all matrices as in \eqref{eq:Proposition_stealthySet}. By the S-procedure~\citep{boyd_linear_1994}, the conditions in~\eqref{eq:DI_Ve_stealthy}--\eqref{eq:DI_InputSet_stealthy} are implied if there exist multipliers ${\alpha}_e, {\beta}_{e,\omega}, {\beta}_{e,v} \in \mathbb{R}_{\geq0}$ such that 
\begin{multline} \label{eq:DI_errorSet_steatlhy}
    -\mathcal M_e - \alpha_e \mathcal N_e - \beta_{e,\omega} \mathcal S_{\kappa,\omega} - \beta_{e,v} \mathcal S_{\kappa,v}  \succeq 0 
\end{multline}  
which is the first LMI in~\eqref{eq:Proposition_stealthySet1}. The residual ellipse $r^\top \Pi r \leq 1$ can be reformulated using \eqref{eq:DI_errorDyn_attFree} to obtain $(Ce + Hv)^\top \Pi (Ce + Hv) \leq 1$. Consequently, $\mathcal{E}_r$ yields
\begin{align} \label{eq:DI_residual_kappa}
    \mathcal E_r &\coloneqq \{r \in \mathbb{R}^{n_y}   \mid \kappa^\top \mathcal M_{r} \kappa \leq 0, && \forall e,\omega,v\},
\end{align}
with $\mathcal{M}_r$ according \eqref{eq:Proposition_stealthySet4}. The set $\mathcal{E}_r$ is nonempty under the assumptions $ e \in \mathcal{E}_e $ and $v \in \mathcal{E}_v$. By the $S$-procedure, a sufficient condition for \eqref{eq:DI_residual_kappa} to hold is the existence of multipliers \( \alpha_r, \beta_{r,v} \in \mathbb{R}_{\geq 0} \) such that  
\begin{align} \label{eq:DI_residualSet_steatlhy}
    -\mathcal M_{r} + \alpha_r \mathcal N_e - \beta_{r,v} \mathcal S_{\kappa,v}  & \succeq 0,
\end{align} 
which concludes the last LMI in~\eqref{eq:Proposition_stealthySet1}.

The joint minimization of $-\log\det(P_e)$ and $-\log\det(\Pi)$ ensures minimal ellipsoidal volumes for the error set $\mathcal{E}_e$ and the residual set $\mathcal{E}_r$, respectively~\citep{murguia_security_2020}; the grid search over $\alpha_e$ and $\alpha_r$ preserves convexity of the search. 
\hfill $\blacksquare$
\end{pf}

\section{Proof of Lemma~\ref{lemma:ErrorSet}} \label{app:proofDetectorEllipsoid1}   
This appendix provides the proof of Lemma~\ref{lemma:ErrorSet}. Define the Lyapunov function candidate 
\begin{align} \label{eq:DI_Ve_att} 
    \bar V(\bar e) = \bar e^\top \bar{P}_e \bar e \geq 0, \quad \bar P_e \succ 0.
\end{align}
For the ellipsoidal set $ \bar{\mathcal{E}}_e$ in \eqref{eq:CR_errorBarEllipse} to be a forward invariant set, we require that~\citep{escudero_analysis_2022,boyd_linear_1994}
\begin{subequations} \label{eq:DI_dVe_att1} \begin{align} 
    \dot{\bar{V}}(\bar e, \omega, v, r) = \dot{e}^\top  \bar{P}_e e + e^\top \bar{P}_e \dot{e} \leq 0,
\end{align}
for all $\bar{e}$ and admissible inputs $\omega \in \mathcal{E}_w, v\in \mathcal{E}_v,$ and $r\in\mathcal{E}_r$, such that $\bar{V}(\bar e) \geq 1$. Substituting~\eqref{eq:DI_ErrorDyn_residue} yields
\begin{multline} \label{eq:DI_dVe_att2} 
    \dot{\bar V}(\bar e, \omega, v, r) = \left(\bar A_e \bar e + \bar G\omega  
     - \bar L(I - \Gamma \Gamma^\dagger)H v \right. \\ \left. - \bar L\Gamma \Gamma^\dagger r \right)^\top \bar{P}_e \bar e 
    + \bar e^\top \bar{P}_e \left( \bar A_e \bar e + \bar G\omega \right. \\ 
    \left. - \bar L(I - \Gamma \Gamma^\dagger)H v - \bar L\Gamma \Gamma^\dagger r \right) \leq 0.
\end{multline}
\end{subequations}
Define the stacked vector $\bar{\kappa} = \begin{bmatrix} \bar e & \omega & v & r & 1 \end{bmatrix}^\top$, which allows the reformulation of~\eqref{eq:DI_Ve_att}, \eqref{eq:DI_dVe_att1}, $\mathcal{E}_\omega$, $\mathcal{E}_v$, and $\mathcal{E}_r$ as 
\begin{subequations} \label{eq:DI_InputSet_lifted} \begin{align}
    \bar V_\kappa( \bar \kappa) &= \bar \kappa^\top \bar{\mathcal{N}}_e \bar \kappa, \quad
    \dot{\bar V}_\kappa(\bar \kappa) = \bar \kappa^\top \bar{\mathcal{M}}_e \bar \kappa, \\
    \! \! \! \mathcal{E}_\omega & \coloneqq \! \{\omega \in \mathbb{R}^{n_x} \! \mid \! \bar{\kappa}^\top \mathcal{S}_{\omega,\bar\kappa} \bar{\kappa} \geq 0,  &&\!\!\!\! \forall \bar e,v,r\}, \\
    \! \! \!  \mathcal{E}_v &\coloneqq \! \{v \in \mathbb{R}^{n_v} \! \mid \! \bar{\kappa}^\top \mathcal{S}_{v,\bar{\kappa}} \bar{\kappa} \geq 0, &&\!\!\!\!\forall \bar e,\omega,r \}, \\
    \! \! \!  \mathcal{E}_r &\coloneqq \! \{r \in \mathbb{R}^{n_y} \! \mid \! \bar{\kappa}^\top \mathcal{S}_{r,\bar\kappa} \bar{\kappa} \geq 0,  &&\!\!\!\!\forall \bar e,\omega,v \}
\end{align} \end{subequations}
with all matrices as in \eqref{eq:lemma_ErrorSet}. By the S-procedure~\citep{boyd_linear_1994}, the inequalities in~\eqref{eq:DI_Ve_att}-\eqref{eq:DI_dVe_att1}, for all  $\omega \in \mathcal{E}_\omega$, $v \in \mathcal{E}_v$, and $r \in \mathcal{E}_r$, are implied if there exists multipliers $\bar{\alpha}_e, \bar{\beta}_{e,\omega}, \bar{\beta}_{e,v}, \bar{\beta}_{e,r} \in \mathbb{R}_{\geq 0}$ such that 
\begin{multline}
    -\bar{\mathcal M}_{e} - \bar{\alpha}_{e} \bar{\mathcal N}_{e} - \bar{\beta}_{e,\omega} \mathcal{S}_{\omega, \bar\kappa} \\ 
            - \bar{\beta}_{e,v} \mathcal{S}_{v, \bar\kappa} - \bar{\beta}_{e,r} \mathcal{S}_{r,\bar\kappa}  \succeq 0. 
\end{multline}  
%with matrices $\bar{\mathcal M}_{e},\bar{\mathcal N}_{e},\mathcal{S}_{\omega, \bar\kappa},\mathcal{S}_{v, \bar\kappa},\mathcal{S}_{r,\bar\kappa}$ defined in~\eqref{eq:lemma_ErrorSet}. 
Since only the subspace $\bar{e}_1$ affects the $\zeta$ dynamics~\eqref{eq:DI_ErrorDyn_residue}, we aim to minimize the projection of $\bar{\mathcal{E}}_e$ onto the subspace $\bar{e}_1$. Let $\bar{e}_1 = \bar{C}_1 \bar{e}$ with $\bar{C}_1 = [I_{n_y} \ 0]$, then the projection of $\bar{\mathcal{E}}_e$ onto $\bar{e}_1$ can be defined as $\bar{\mathcal{E}}_{e_1} \coloneqq \{ \bar{e}_1 \in \mathbb{R}^{n_y} \mid 1 - \bar{e}_1^\top (\bar{C}_1 \bar{P}_e^{-1} \bar{C}_1^\top)^{-1} \bar{e}_1 \geq 0 \}$. By introducing positive definite matrix $Z \in \mathbb{R}^{n_y \times n_y}$ and applying the Schur complement, we obtain the inequality in~\eqref{eq:lemma_ErrorSet1}~~\citep{boyd_linear_1994}:
    \begin{equation}
        \begin{bmatrix} Z & \bar{C}_1 \\ \bar{C}_1^\top & \bar{P}_e \end{bmatrix} \succeq 0 \iff Z \succeq \bar{C}_1 \bar{P}_e^{-1} \bar{C}_1^\top.
    \end{equation}
Thus, by minimizing $\text{trace}(Z)$, we minimize an upper bound on the sum of the semi-axes of $\bar{\mathcal{E}}_{e_1}$. Feasibility of the convex program in the lemma implies that the ellipsoid $\bar{\mathcal{E}}_e$ is forward-invariant and contains the infinite-time reachable set $\bar{\mathcal{R}}^\infty_e$. The problem is solved via a grid search over $\bar \alpha_e$ to preserve convexity.
%\end{pf}

%%%%%%%%%%%%%%%%%%%%%%%%%%%%%%%%%%%%%%%%%%%%%%%%%%%%%%%%%%%%%%%%%%%%%%%%%%%%%%%%
% AUTHOR BIOGRAPHIES
%%%%%%%%%%%%%%%%%%%%%%%%%%%%%%%%%%%%%%%%%%%%%%%%%%%%%%%%%%%%%%%%%%%%%%%%%%%%%%%%

% --- Author 1 ---
\setlength{\columnsep}{7.5pt}
\begin{wrapfigure}{l}{26mm}
  \vspace{-10pt}
  \includegraphics[width=25mm,height=32mm,clip,keepaspectratio]{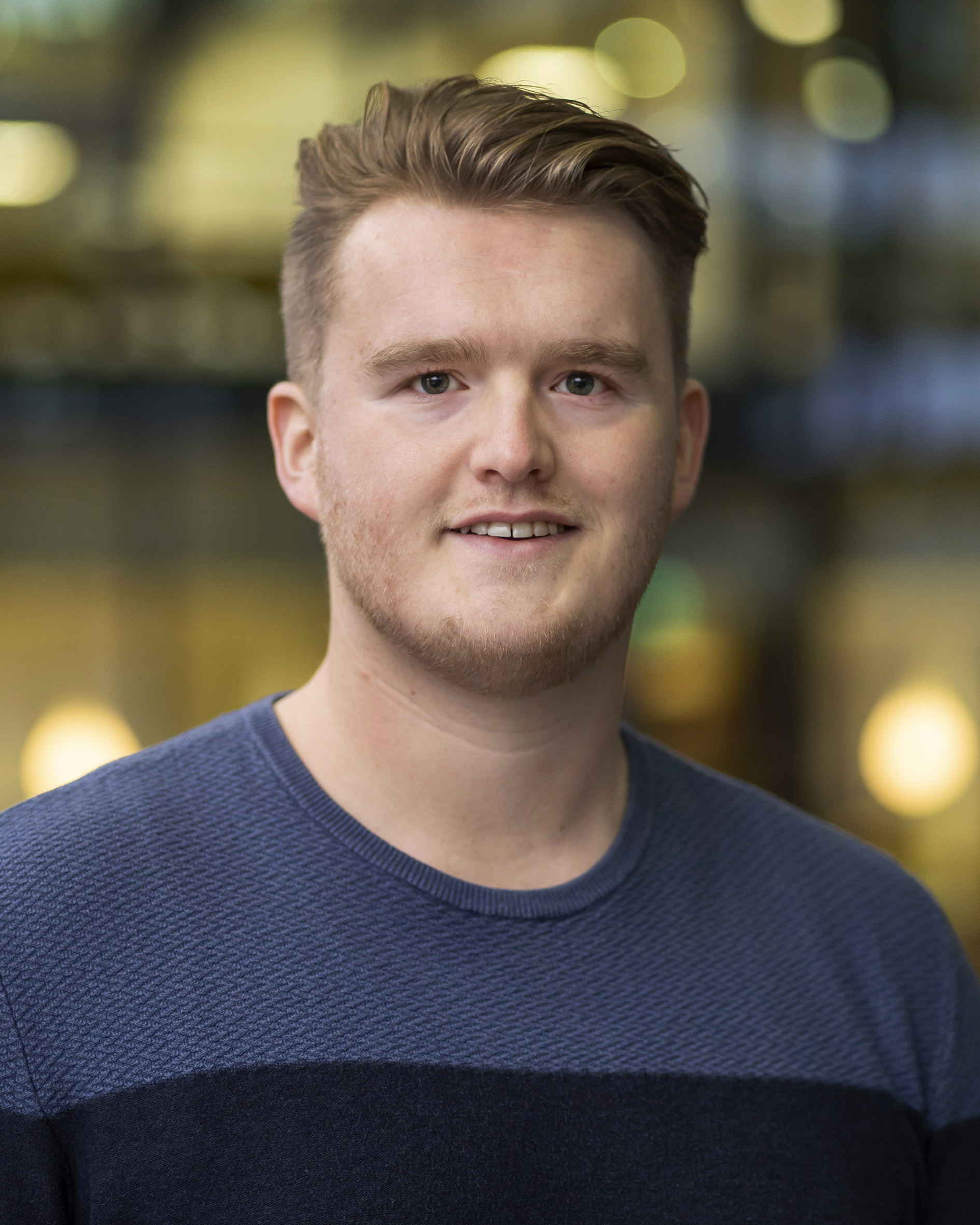}
  \vspace{-25pt}
\end{wrapfigure}
\noindent
\textbf{Mischa Huisman} received the double B.Sc. degree in mechanical engineering from Windesheim University of Applied Sciences, Zwolle, the Netherlands, and Linnaeus University, V{\"a}xj{\"o}, Sweden, in 2018, and the M.Sc. degree in automotive technology from Eindhoven University of Technology, Eindhoven, the Netherlands, in 2022. He is currently pursuing the Ph.D. degree with the Dynamics and Control Group, Department of Mechanical Engineering, Eindhoven University of Technology. His research interests include resilient control for cyber-physical systems, with a focus on mitigating cyberattacks, parametric uncertainty, and model mismatch. His work applies robust and secure control techniques to cooperative driving, vehicle platooning, and cooperative adaptive cruise control.
\vspace{1.5em}

% --- Author 2 ---
\begin{wrapfigure}{l}{26mm}
  \vspace{-10pt}
  \includegraphics[width=25mm,height=32mm,clip,keepaspectratio]{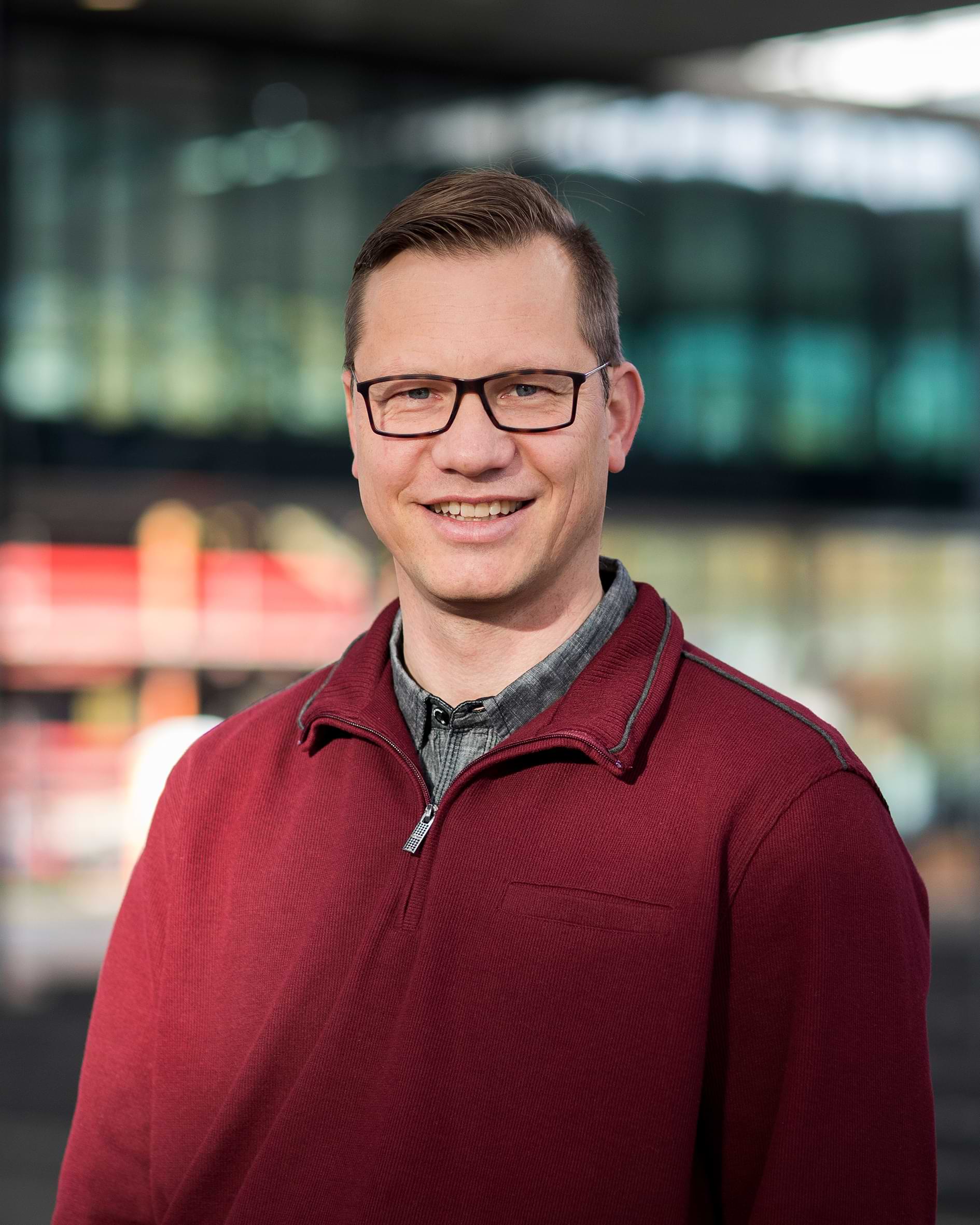}
  \vspace{-4pt}
\end{wrapfigure}
\noindent
\textbf{Erjen Lefeber} Erjen Lefeber received the M.Sc. and Ph.D. degrees in Nonlinear Control from the Department of Applied Mathematics at the University of Twente, The Netherlands, in 1996 and 2000, respectively. He is an Assistant Professor with the Department of Mechanical Engineering at Eindhoven University of Technology, Eindhoven, The Netherlands, where he is a member of the Dynamics and Control group. His research interests include nonlinear control of mechanical systems, with applications to cooperative and automated road vehicles, including passenger cars and trucks, as well as underactuated aerial vehicles. His work covers problems such as vehicle platooning, cooperative adaptive cruise control, truck control, and quadcopter trajectory tracking, with an emphasis on stability, robustness, security, safety, and performance under communication constraints and uncertainties.
\vspace{1.5em}

% --- Author 3 ---
\begin{wrapfigure}{l}{26mm}
  \vspace{-10pt}
  \includegraphics[width=25mm,height=32mm,clip,keepaspectratio]{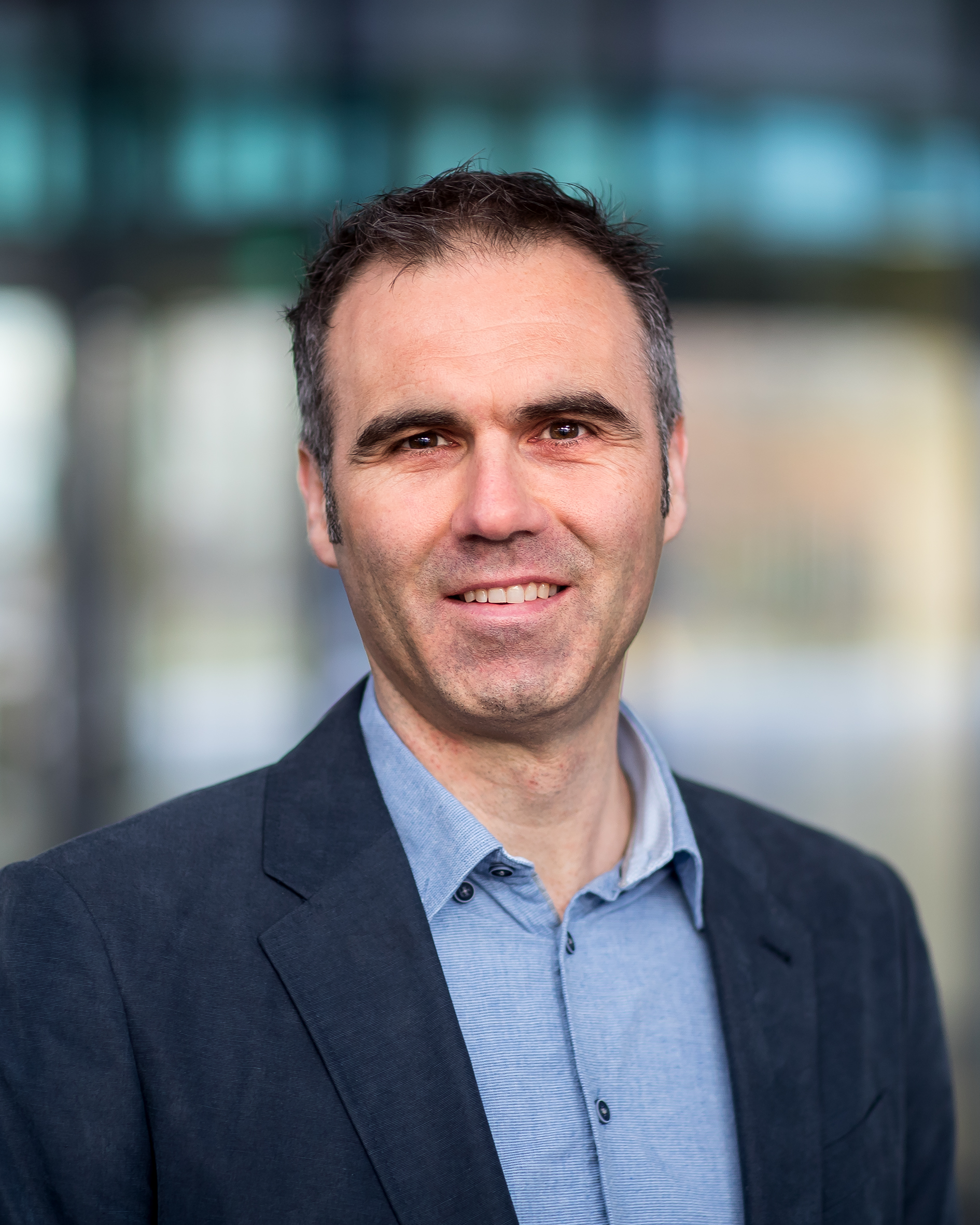}
  \vspace{-4pt}
\end{wrapfigure} 
\noindent
\textbf{Nathan van de Wouw} obtained his M.Sc.-degree (with honours) and Ph.D.-degree in Mechanical Engineering from the Eindhoven University of Technology, the Netherlands, in 1994 and 1999, respectively. He currently holds a full professor position at the Mechanical Engineering Department of the Eindhoven University of Technology, the Netherlands. He has been working at Philips Applied Technologies, The Netherlands, in 2000 and at the Netherlands Organisation for Applied Scientific Research, The Netherlands, in 2001. He has been a visiting professor at the University of California Santa Barbara, U.S.A., in 2006/2007, at the University of Melbourne, Australia, in 2009/2010 and at the University of Minnesota, U.S.A., in 2012 and 2013. He has held a (part-time) full professor position the Delft University of Technology, the Netherlands, from 2015-2019. He has also held an adjunct full professor position at the University of Minnesota, U.S.A, from 2014-2021. He has published the books 'Uniform Output Regulation of Nonlinear Systems: A convergent Dynamics Approach' with A.V. Pavlov and H. Nijmeijer (Birkhauser, 2005) and `Stability and Convergence of Mechanical Systems with Unilateral Constraints' with R.I. Leine (Springer-Verlag, 2008). In 2015, he received the IEEE Control Systems Technology Award "For the development and application of variable-gain control techniques for high-performance motion systems". He also received the Springer Nature Award for the Best Paper in Nonlinear Dynamics in 2025 and the 2026 Best Application Paper Award of the IFAC WC. He is an IEEE Fellow for his contributions to hybrid, data-based and networked control.
\vspace{1.5em}

% --- Author 4 ---
\begin{wrapfigure}{l}{26mm}
  \vspace{-10pt}
  \includegraphics[width=25mm,height=32mm,clip,keepaspectratio]{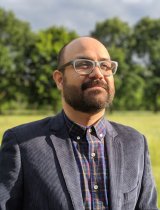}
  \vspace{-4pt}
\end{wrapfigure}
\noindent
\textbf{Carlos Murguia} received his Ph.D. degree in Mechanical Engineering from Eindhoven University of Technology (TU/e), The Netherlands, in 2015. He is currently an Associate Professor in the Engineering Systems and Design Pillar at the Singapore University of Technology and Design (SUTD), Singapore, and an adjunct faculty member (part-time) in the Dynamics and Control Group at TU/e. From 2020 to 2024, he was a full-time Assistant Professor at TU/e, where he received tenure in 2023. Between 2015 and 2020, he held research positions at the Centre for Research in Cyber Security (iTrust) at SUTD, the University of Melbourne, Princeton University, and the University of California, Los Angeles. He was also a Visiting Faculty member at Queensland University of Technology from 2023 to 2024. He has served as principal investigator, subproject leader, or technical task leader on competitive research projects funded by Horizon Europe, the Dutch Research Council, the Singapore Ministry of Education, and TU/e. These include the Horizon Europe SELFY and SecReSy4You projects, the NWO DIGITAL TWIN and N-HIPO projects, and Singapore Ministry of Education Tier 1 and Tier 2 projects. He has authored more than 80 journal and conference papers in top systems and control venues and currently serves as an Associate Editor for the IEEE Transactions on Control of Network Systems and the International Journal of Robust and Nonlinear Control. His research develops resilient cyber-physical systems that maintain safety and performance despite faults, cyberattacks, uncertainty, and changing operating conditions. His work combines systems and control theory with physics-based modelling, data-driven methods, and artificial intelligence to advance fault and attack diagnosis, secure and privacy-preserving control, and resilient control with formal guarantees. Applications include advanced manufacturing, robotics, autonomous mobility, and critical infrastructure.

\end{document}